\documentclass[aps,prd,10pt,preprintnumbers,nofootinbib,superscriptaddress,letterpaper,twocolumn]{revtex4-1}
\pdfoutput=1
\usepackage{amsmath,amssymb,amsbsy,multirow}
\usepackage[utf8]{inputenc}
\usepackage{graphicx}
\usepackage[colorlinks=true, linkcolor=blue, filecolor=blue, urlcolor=blue, citecolor=blue, pdftex, plainpages=false]{hyperref}

\newcommand{\eq}[1]{Eq.~(\ref{#1})}

\begin{document}
%----------------------------------------------------------------------------
\title{Near-conformal dynamics in a chirally broken  system}
\author{Thomas~Appelquist}
\affiliation{Department of Physics, Sloane Laboratory, Yale University, New Haven, CT 06520, United States}
\author{Richard~C.~Brower}
\affiliation{Department of Physics and Center for Computational Science, Boston University, Boston, MA 02215, United States}
\author{Kimmy~K.~Cushman}
\affiliation{Department of Physics, Sloane Laboratory, Yale University, New Haven, CT 06520, United States}
\author{George~T.~Fleming}
\affiliation{Department of Physics, Sloane Laboratory, Yale University, New Haven, CT 06520, United States}
\author{Andrew~D.~Gasbarro}
\affiliation{Institute for Theoretical Physics, University of Bern, 3012 Bern, Switzerland}
\author{Anna~Hasenfratz}
\affiliation{Department of Physics, University of Colorado, Boulder, CO 80309, United States}
\author{Xiao-Yong~Jin}
\affiliation{Leadership Computing Facility, Argonne National Laboratory, Argonne, IL 60439, United States}
\author{Ethan T.~Neil}
\affiliation{Department of Physics, University of Colorado, Boulder, CO 80309, United States}
\author{James C.~Osborn}
\affiliation{Leadership Computing Facility, Argonne National Laboratory, Argonne, IL 60439, United States}
\author{Claudio~Rebbi}
\affiliation{Department of Physics and Center for Computational Science, Boston University, Boston, MA 02215, United States}
\author{Enrico~Rinaldi}
\affiliation{Arithmer Inc., R\&D Headquarters, Minato, Tokyo 106-6040, Japan and Interdisciplinary Theoretical and\\  Mathematical Sciences Program (iTHEMS), RIKEN,  2-1 Hirosawa, Wako, Saitama 351-0198, Japan}
\author{David~Schaich}
\affiliation{Department of Mathematical Sciences, University of Liverpool, Liverpool L69 7ZL, United Kingdom}
\author{Pavlos~Vranas}
\affiliation{Nuclear and Chemical Sciences Division, Lawrence Livermore National Laboratory, Livermore, CA 94550, United States}
\affiliation{Nuclear Science Division, Lawrence Berkeley National Laboratory, Berkeley, CA 94720, United States}
\author{Oliver~Witzel}
\email[Corresponding author: ]{oliver.witzel@uni-siegen.de}
\altaffiliation[present address: ]{Theoretische Physik 1, Naturwissenschaftlich-Technische Fakult\"at, Universit\"at Siegen, 57068 Siegen, Germany}
\affiliation{Department of Physics, University of Colorado, Boulder, CO 80309, United States}

\collaboration{Lattice Strong Dynamics Collaboration}

\preprint{LLNL-JRNL-812164}

\date{\today}
\begin{abstract}
Composite Higgs models must exhibit very different dynamics from quantum chromodynamics (QCD) regardless whether they describe the Higgs boson as a dilatonlike state or a pseudo-Nambu-Goldstone boson. Large separation of scales and large anomalous dimensions are frequently  desired by phenomenological models. Mass-split systems are well-suited for composite Higgs models because they are governed by a conformal fixed point in the ultraviolet but are chirally broken in the infrared.  In this work we use lattice field theory calculations with domain wall fermions to investigate a system with four light and six heavy flavors. We demonstrate how a nearby conformal fixed point  affects the  properties of the four light flavors that exhibit chiral symmetry breaking in the infrared.  Specifically we describe hyperscaling of dimensionful physical quantities and determine the corresponding anomalous mass dimension. We obtain $y_m=1+\gamma^*= 1.47(5)$ suggesting that $N_f=10$ lies inside the conformal window. Comparing the low energy spectrum to predictions of dilaton chiral perturbation theory, we observe excellent agreement which supports the expectation that the 4+6 mass-split system exhibits near-conformal dynamics with a relatively light $0^{++}$ isosinglet scalar.
\end{abstract}

\maketitle
%----------------------------------------------------------------------------
\section{Introduction}
Experiments have discovered a $125$ GeV Higgs boson \cite{Aad:2012tfa,Chatrchyan:2012ufa,Aad:2015zhl} but so far, up to a few TeV, no direct signs of physics beyond the standard model (BSM) have been seen.  The standard model (SM), however, is an effective theory, and new interactions are necessary to UV complete the Higgs sector, explain dark matter, and account for the matter-antimatter asymmetry of the universe. For gauge theories describing the Higgs sector as a composite structure,  experimental observations imply that a large separation of scales between the electroweak scale (IR) and new ultraviolet physics (UV) \cite{Contino:2010rs,Luty:2004ye,Dietrich:2006cm,Luty:2008vs,Brower:2015owo,Csaki:2015hcd,Arkani-Hamed:2016kpz,Witzel:2019jbe} is required. Theories with a large separation of scales part company from QCD, exhibiting a ``walking'' gauge coupling \cite{Yamawaki:1985zg,Appelquist:1986an,Bando:1987br}, and providing a dynamical mechanism for electroweak (EW) symmetry breaking. They can satisfy stringent constraints from EW precision measurements but avoid unnaturally large tuning of the Higgs mass.

To explore theories with a large scale separation and infrared dynamics different from QCD, we employ the device of mass splitting \cite{Brower:2014dfa,Brower:2015owo,Hasenfratz:2016gut,Hasenfratz:2017hdd}, where the action has two fermion mass parameters: $\widehat m_h$ and a smaller $\widehat m_\ell$. The idea is to start with sufficiently many fermions to guarantee that at scales well above the fermion masses the theory exhibits infrared conformality.  By assigning the masses $\widehat m_h$ and $\widehat m_{\ell}$ to the fermions, we create a system with $N_h$ heavy fermions and $N_{\ell}$ light fermions. The number of light fermions $N_{\ell}$ is chosen such that the light sector alone exhibits spontaneous chiral symmetry breaking. The resulting mass-split theory is governed by the conformal IRFP above the high scale set by $\widehat m_h$.  There the spectrum exhibits conformal hyperscaling, and the mass of the lightest isosinglet scalar $0^{++}$ is expected to be comparable to the corresponding pseudoscalar mass \cite{Miransky:1998dh,Aoki:2013zsa}.

In the infrared, the heavy fermions decouple, the gauge coupling runs to larger values, and chiral symmetry for the light flavors breaks spontaneously.  The heavy-fermion mass $\widehat m_h$  controls the separation of scales between the UV and IR  \cite{Hasenfratz:2017hdd}. Even though the low energy theory is chirally broken, its properties are significantly different from a QCD-like theory with $N_\ell$ fermions. In particular a light $0^{++}$ state may enter the effective chiral Lagrangian, requiring the extension to dilaton chiral perturbation theory (dChPT) \cite{Golterman:2016lsd,Appelquist:2017vyy,Appelquist:2017wcg,Golterman:2018mfm,Appelquist:2019lgk,Golterman:2020tdq}.

 It is favorable to keep the total number of fermions $N_f = N_h +N_\ell$ near the low end of the conformal window to achieve a large anomalous dimension. Specifically we study an $SU(3)$ gauge theory with four light and six heavy fermions in the fundamental representation. Although no consensus has been reached on the precise onset of the conformal window for $SU(3)$ gauge theories with fundamental fermions, there are indications that $N_f= 10$ is infrared conformal \cite{Hayakawa:2010yn,Appelquist:2012nz,Chiu:2016uui,Chiu:2017kza,Chiu:2018edw,Fodor:2017gtj,Hasenfratz:2017qyr,Hasenfratz:2020ess,Baikov:2016tgj,Ryttov:2010iz,Ryttov:2016ner,Ryttov:2016hal,Ryttov:2017kmx,Antipin:2018asc,DiPietro:2020jne}. By choosing a theory with four fermions in the light, chirally broken sector, our simulations can also directly be related to existing models extending the SM with a new strongly interacting sector \cite{Ma:2015gra,BuarqueFranzosi:2018eaj,Marzocca:2018wcf}.   In these models the Higgs boson is a pseudo-Nambu-Goldstone boson (pNGB) of the new strong sector \cite{Vecchi:2015fma,Ferretti:2013kya,Ferretti:2016upr,Ma:2015gra,BuarqueFranzosi:2018eaj}.

We explore this new, strongly coupled theory by performing large scale numerical lattice-field-theory simulations. The choice $N_f= 10$ improves over a pilot study using four light and eight heavy flavors \cite{Brower:2014dfa,Brower:2015owo,Hasenfratz:2016gut,Brower:2014ita,Weinberg:2014ega,Hasenfratz:2015xca,Hasenfratz:2016uar,Hasenfratz:2017lne}  by being closer to the bottom of the conformal window. Also, we perform the numerical simulations using chiral domain-wall fermions (DWF)   \cite{Kaplan:1992bt,Shamir:1993zy,Furman:1994ky,Brower:2012vk} which preserve the flavor structure. While numerically more costly, DWF provide a theoretically clean environment to perform investigations of strongly coupled theories near a conformal IR fixed point.

We briefly introduce the details of the numerical simulations before we demonstrate hyperscaling and determine the mass anomalous dimension. This allows us to explore implications for a possible effective description at low energies. Finally we give an outlook on our future calculations of phenomenologically important quantities. Preliminary results have been reported in \cite{Witzel:2018gxm,Witzel:2019oej}.

\section{Numerical simulations}
Simulations are performed on hypercubic lattices using $(L/a)^3 \times T/a$ volumes with $L/a=24$ or 32 and $T/a=64$ where $a$ indicates the lattice spacing. 
We simulate the $SU(3)$ gauge system with four light and six heavy flavors using the Symanzik gauge action \cite{Luscher:1984xn,Luscher:1985zq} with 3-times stout-smeared ($\rho=0.1$) \cite{Morningstar:2003gk} M\"obius domain wall fermions \cite{Brower:2012vk} ($b_5=1.5$, $c_5=0.5$). DWF are simulated by adding a fifth dimension of extent $L_s$ which separates the physical modes of four dimensional space-time. For practical reasons $L_s$ needs to be finite i.e.~DWF exhibit a small, residual chiral symmetry breaking, conventionally parametrized as an additive mass term $am_\text{res}$. In our simulations we choose $L_s=16$ and set the domain wall height $M_5=1$. We determine the residual chiral symmetry breaking numerically and find small values of $O({10}^{-3})$. To correctly refer to the dimensionless lattice masses, we introduce the notation
\begin{align}
\widehat m_x \equiv  a \widetilde m_x = a (m_x + m_\text{res}) \quad\text{with}\quad x=\ell,\,h.
\end{align}

\begin{figure}[tb]
  \centering
  \includegraphics[width=\columnwidth]{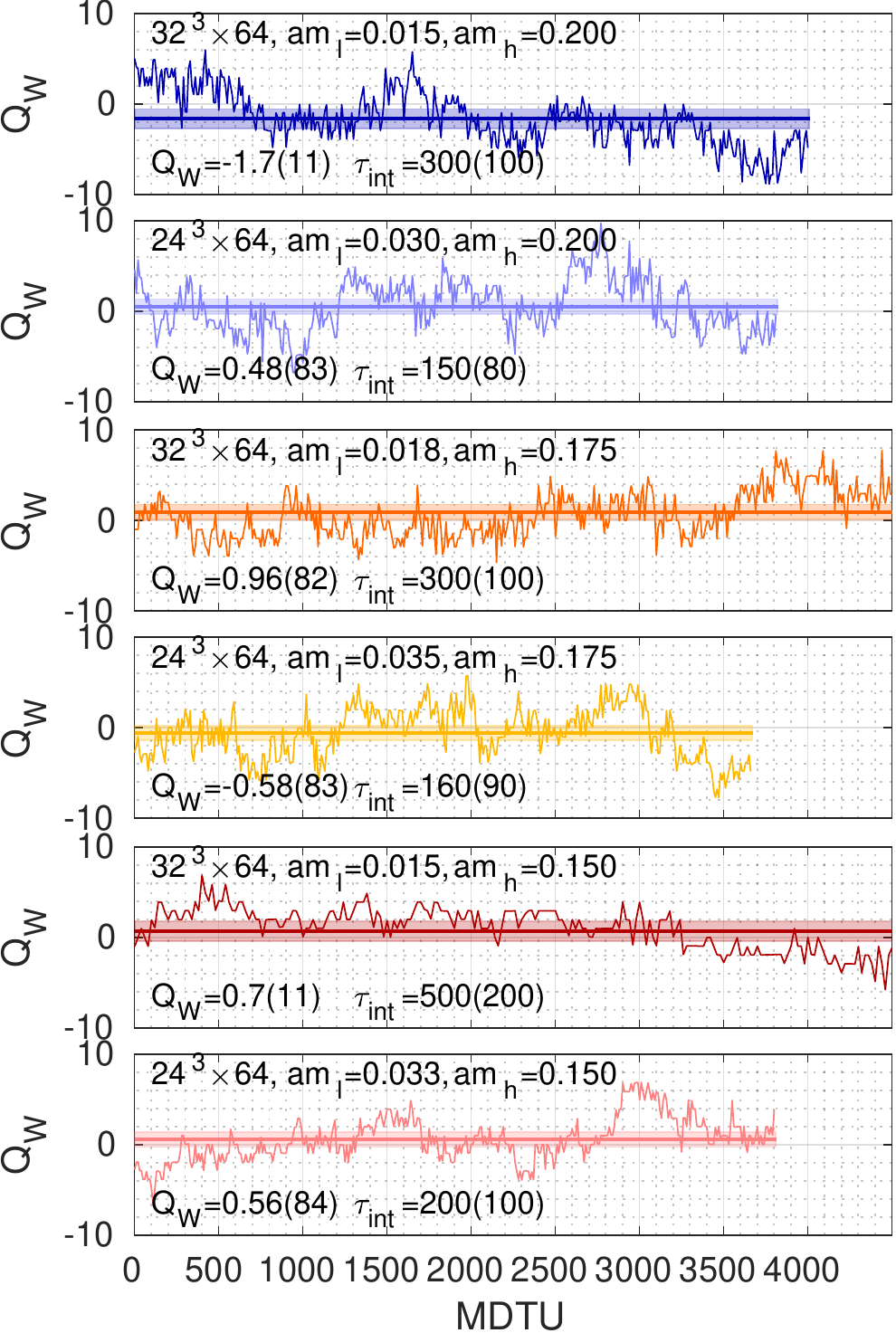}
  \caption{Monte Carlo histories of the topological charge obtained using Wilson gradient flow at flow time $t=L^2/32$ for six high statistics ensembles. The integrated autocorrelation time $\tau_\text{int}$ of the topological charge $Q_W$ is determined using the $\Gamma$-method \cite{Wolff:2003sm} and quoted in units of MDTU. }
  \label{Fig.topoHistory}
\end{figure}

Based on insight from our accompanying step-scaling investigation \cite{Hasenfratz:2017qyr,Hasenfratz:2020ess}, we set the bare gauge coupling $\beta=6/g_0^2=4.03$, close to the IRFP of the underlying conformal theory with ten degenerate flavors.  
The hybrid Monte Carlo (HMC) update algorithm \cite{Duane:1987de} with a trajectory length of $\tau=2$ MDTU (molecular dynamics time units) is used to generate ensembles of dynamical gauge field configurations  with $1-3$k ($0.3-0.5$k) thermalized trajectories for $am_\ell\le 0.04$ ($am_\ell>0.04$). Using input heavy flavor mass $am_h=0.200$, 0.175, and 0.150, we explore the 4+6 system choosing five or seven values for the input light flavor mass in the range $0.015\le am_\ell\le 0.100$.  Spectrum measurements are performed every 20 (10) MDTU for $am_\ell< 0.04$ ($am_\ell\ge 0.04$) and we decorrelate subsequent measurements by randomly choosing source positions. Remaining autocorrelations are estimated using the $\Gamma$-method \cite{Wolff:2003sm} and accounted for by correspondingly binning subsequent measurements in our jackknife analysis.  For all ensembles we observe frequent changes of the topological charge measured every 10 MDTU (20 MDTU for $m_\ell/m_h=0.015/0.150$) which typically is a quantity exhibiting the longest autocorrelation times in the system. Examples of the Monte Carlo histories for six high statistics ensembles are shown in Fig.~\ref{Fig.topoHistory}.

\section{Hyperscaling}

To understand the properties of mass-split systems, we refer to Wilsonian renormalization group (RG).
In the UV  both mass parameters are much lighter than the cutoff $\Lambda_\text{cut}=1/a$: $\widehat m_l \ll 1$, $\widehat m_h\ll 1$.
As the energy scale $\mu$ is lowered from the cutoff, the RG flowed lattice action moves in the infinite parameter action space as dictated by the fixed point structure of the $N_f$ flavor conformal theory. The masses are increasing according to their scaling dimension $y_m$, $\widehat m_{\ell,h} \to \widehat m_{\ell,h} (a \mu)^{-y_m}$, but we assume that they are still small so the system remains close to the conformal critical surface. All other couplings are irrelevant and approach the IRFP as the energy scale is lowered.

If the gauge couplings reach the vicinity of the IRFP, only the two masses change under RG flow. We can use standard hyperscaling arguments \cite{DeGrand:2009mt,DelDebbio:2010ze,DelDebbio:2010jy} to show that any physical quantity $aM_H$ of mass dimension one follows, at leading order, the scaling form \cite{Hasenfratz:2016gut} 
 \begin{align}
 a M_H = \widehat m_h^{1/y_m}  \Phi_H( \widehat m_\ell/ \widehat m_h), 
\label{eq:M_scaling} 
 \end{align}
 where $y_m=1+\gamma_m^\star$ is the universal scaling dimension of the mass at the IRFP and $\Phi_H$  some function of $ \widehat m_\ell/ \widehat m_h$. $\Phi_H$ depends on the observable $H$ and could be qualitatively different for different $H$.\footnote{Equivalent to Eq.~(\ref{eq:M_scaling}) is the hyperscaling relation, $aM_H= \widehat m_l^{1/y_m}  \Phi_H( \widehat m_\ell/ \widehat m_h)$, given in Ref.~\cite{Hasenfratz:2016gut}. Depending on the observable and scaling test, one or the other form might be preferable.}
The scaling relation Eq.~(\ref{eq:M_scaling}) is valid as long as the gauge couplings remain at the IRFP and lattice masses are small, i.e.~even in the $\widehat m_\ell =0$ chiral limit.
 As a consequence, ratios of masses
  \begin{align}
\frac{M_{H1} }{M_{H2}} = \frac{ \Phi_{H1}( \widehat m_\ell/  \widehat m_h)}{  \Phi_{H2}( \widehat m_\ell/ \widehat m_h)} 
\label{eq:R_scaling}
 \end{align}
depend only on $\widehat m_\ell/ \widehat m_h$. 
 The heavy flavors decouple when $\widehat m_h (a\mu)^{-y_m} \approx 1$. At that point the light flavors condense and spontaneously break chiral symmetry. This allows us to define the hadronic or chiral symmetry breaking scale 
 \begin{align}
 \Lambda_H = \widehat m_h^{1/y_m}  a^{-1}.
\label{eq:Lambda_H}
 \end{align} 
As the energy scale $\mu$ is lowered below $\Lambda_H$, the gauge coupling starts running again. However, properties of the IRFP are already encoded in hadronic observables. 
\begin{figure}[tb]
  \centering
  \includegraphics[width=\columnwidth]{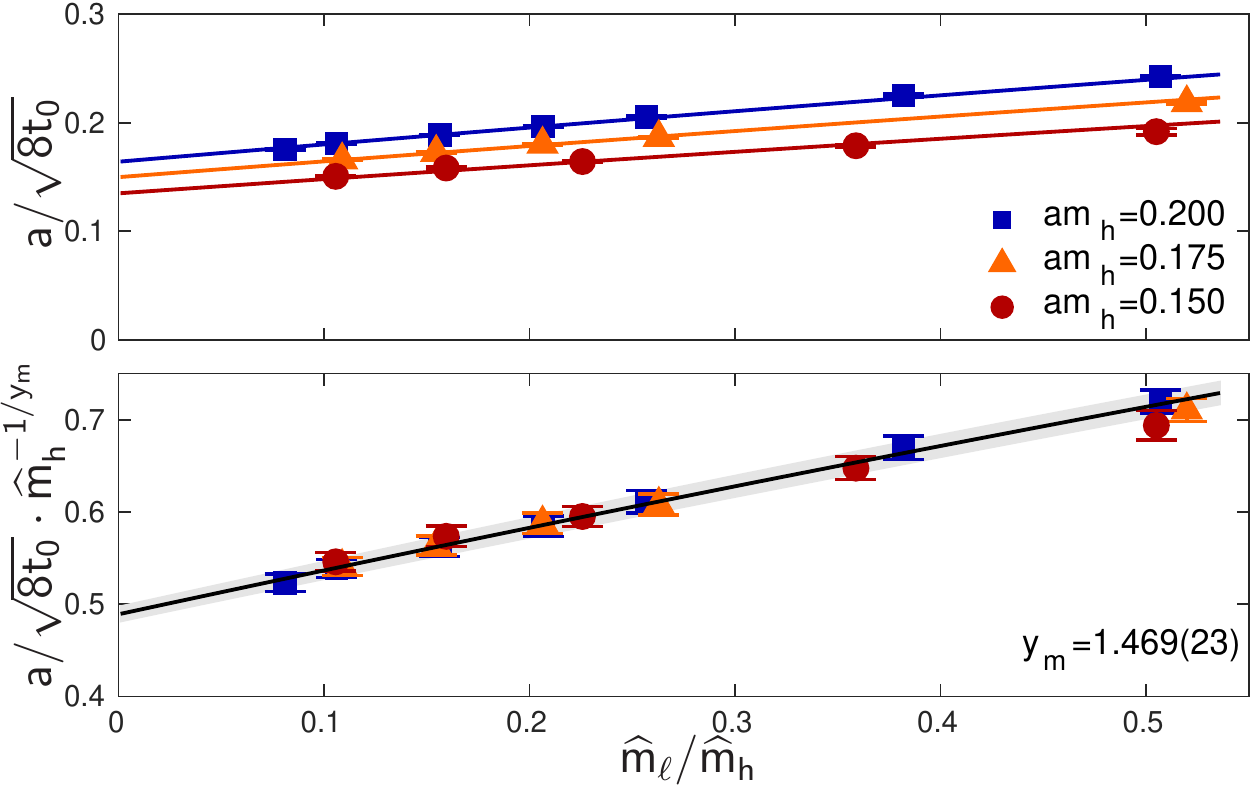}
 \caption{The inverse  Wilson flow scale $a/\sqrt{8t_0}$ and interpolating fit according to \eq{eq:M_scaling} as function of $\widehat m_\ell/\widehat m_h$. The bottom panel shows ``curve collapse''  for $\Phi_{\sqrt{8t_0}}(\widehat m_\ell/\widehat m_h)= a /\sqrt{8t_0}\cdot \widehat m_h^{- 1/y_m}$.} 
  \label{Fig.hyperscaling_s8t0}
\end{figure}
 We have  established hyperscaling of ratios in the 4+8 flavor system \cite{Brower:2015owo,Hasenfratz:2016gut} and preliminary results for the 4+6 system  are reported in \cite{Witzel:2018gxm,Witzel:2019oej}.

In Fig.~\ref{Fig.hyperscaling_s8t0} we illustrate hyperscaling and the determination of $y_m$ by considering the inverse Wilson flow scale $a/\sqrt{8t_0}$ as the quantity $aM_H$ in Eq.~(\ref{eq:M_scaling}).
 The dimensionful quantity $1/\sqrt{8t_0}$ is proportional to the energy scale where the renormalized running coupling in the gradient flow scheme equals a reference value ($g^2_{GF}\approx 16$) \cite{Luscher:2010iy}. 
 The top panel shows $a/\sqrt{8t_0}$ as the function of $\widehat m_\ell/\widehat m_h$.   While the data corresponding to our three different $am_h$ values are different, each set on its own follows a smooth, almost linear curve. This suggests to parametrize the unknown function $\Phi_{\sqrt{8t_0}}(\widehat m_\ell/\widehat m_h)$  using a low-order polynomial and perform a combined fit to all 17 data points in Fig.~\ref{Fig.hyperscaling_s8t0} using the  Ansatz given in \eq{eq:M_scaling}. 
A fit with a quadratic polynomial describes our data well. Small deviations of very precise $a/\sqrt{8t_0}$ values lead to $\chi^2/\text{d.o.f.}\approx 3$ and $y_m=1.469(23)$ with likely underestimated statistical uncertainties.

\begin{figure}[tb]
  \centering
  \includegraphics[width=\columnwidth]{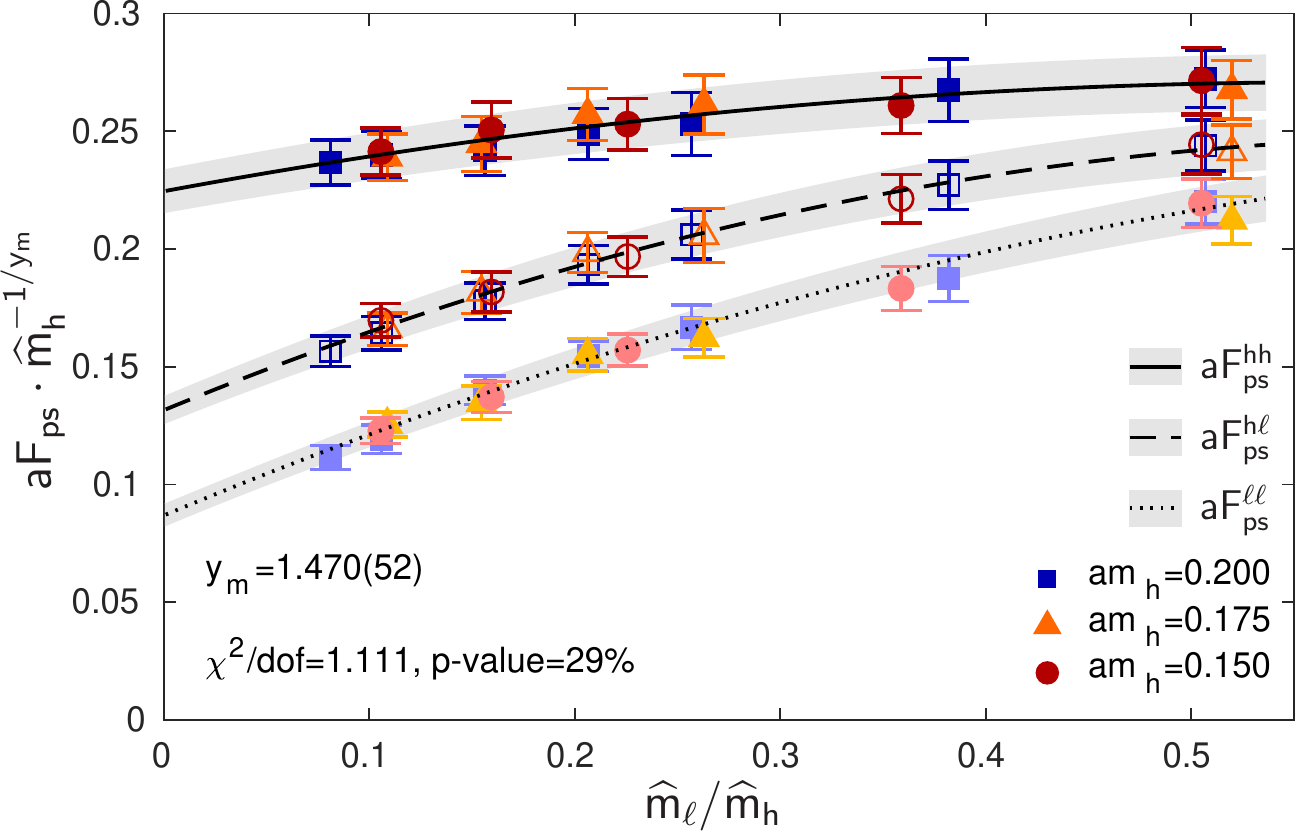}
  \caption{Hyperscaling of the light-light (shaded symbols), heavy-light (open symbols), and heavy-heavy (filled symbols) pseudoscalar decay constant as the function of $\widehat m_\ell/\widehat m_h$. A combined fit based on Eq.~(\ref{eq:M_scaling}) determines $y_m$.}
  \label{Fig.hyperscaling_ps_decay}
\end{figure}

The bottom panel of Fig.~\ref{Fig.hyperscaling_s8t0} shows the data points for $a/\sqrt{8t_0} \cdot \widehat m_h^{-1/y_m}$ and the quadratic fit function $\Phi_{\sqrt{8t_0}}(\widehat m_\ell/\widehat m_h)$, exhibiting the expected ``curve collapse.''
We find similar curve collapse  for other observables and show in Fig.~\ref{Fig.hyperscaling_ps_decay}  the result for a combined, correlated fit to the light-light $(\ell\ell)$, heavy-light $(h\ell)$, and heavy-heavy $(hh)$ pseudoscalar decay constant $aF_{ps}$. Since the determination of $aF_{ps}$ is equally precise for $\ell\ell$, $h\ell$, or $hh$ states, this fit provides a representative determination of $y_m$ with a good $p$-value. Subsequently we use
  \begin{align}
    y_m =  1+ \gamma_m^\star =1.470(52),
    \label{eq:ym}
  \end{align}
  as our reference value and note it is consistent within uncertainties to determinations from other observables like vector or pseudoscalar masses.  Further $y_m$ is in agreement to an independent determination based on gradient flow \cite{AnomDim:2020} and comparable to predictions from analytical calculations \cite{Ryttov:2016asb,Ryttov:2017kmx,Ryttov:2017lkz}. 
The predicted  $\gamma_m^\star$ is substantially below 1, the value expected for a system close to the sill of the conformal window \cite{Yamawaki:1985zg,Matsuzaki:2013eva}. Since dChPT analysis of the $N_f=8$ data \cite{Appelquist:2016viq,Appelquist:2018yqe} predicts $\gamma_m^*$ near 1 \cite{Golterman:2016lsd,Appelquist:2017vyy,Appelquist:2017wcg,Golterman:2018mfm,Appelquist:2019lgk,Golterman:2020tdq}, this indicates the sill of the conformal window lies between $N_f=8$ and 10, whereas the 12 flavor system ($\gamma_m^\star\approx 0.24$ \cite{Appelquist:2011dp,DeGrand:2011cu,Cheng:2013eu,Cheng:2013xha,Lombardo:2014pda,Ryttov:2016asb,Ryttov:2017kmx,Ryttov:2017lkz,Li:2020bnb}) is even deeper in the conformal regime.

 The scaling of $a/\sqrt{8t_0}$  is particularly interesting  because it shows that the lattice spacing in the $\widehat m_\ell=0$ chiral limit has a simple dependence on the heavy flavor mass
 \begin{align}
a =  (\widehat m_h)^{1/y_m} \cdot  \Phi_{\sqrt{8t_0}}(0)  \cdot \sqrt{8 t_0}|_{m_\ell=0},
\label{eq:latticespacing_ch}
 \end{align} 
where $\Phi_{\sqrt{8t_0}}(0)$ is a finite number, $\approx0.48$, in the 4+6 system. This confirms the expectation that the continuum $a=0$ limit is approached as $\widehat m_h$ decreases. Combined with \eq{eq:Lambda_H} it predicts the hadronic scale  
\begin{align}
\Lambda_H^{-1} =    \Phi_{\sqrt{8t_0}}(0)  \cdot \sqrt{8 t_0}|_{m_\ell=0}.
\label{eq:Lambda_H2}
 \end{align}

\section{Low energy effective description}

In the low energy infrared limit our system exhibits spontaneous chiral symmetry breaking. It should be described by a chiral effective Lagrangian which smoothly connects to the hyperscaling relation \eq{eq:M_scaling},  valid at the hadronic scale $\mu=\Lambda_H$. In order to combine data sets with different $\widehat m_h$, we express the lattice scale $a$ in terms of the hadronic scale $\Lambda_H$  
\begin{align}
   M_H/\Lambda_H    = (aM_H)\cdot{\widehat m}_h^{-1/y_m} &=   \Phi_H( \widehat m_\ell/ \widehat m_h). 
\label{eq:M_scaling_LH}
\end{align}
Below the hadronic scale $\Lambda_H$, the 4+6 system reduces to a chirally broken $N_f=4$ system.  The low energy effective theory (EFT) expresses the dependence  of physical quantities on the running fermion mass $m_f$ 
of the light flavors. At the hadronic energy scale  the light flavor mass in lattice units is $\widehat m_\ell (a \Lambda_H)^{-y_m}$, predicting \begin{align}
m_f \propto \widehat m_\ell (a \Lambda_H)^{-y_m}\cdot \Lambda_H  = ( \widehat m_\ell/ \widehat m_h) \cdot \Lambda_H.
\label{eq:m_f}
\end{align}
The continuum limit is taken by tuning $\widehat m_h \to 0$ while keeping $\widehat m_\ell/ \widehat m_h$ fixed.

For $\widehat m_\ell /\widehat m_h \lesssim 1$, we expect the $0^{++}$ ground state to be dominated by the light fermions. It is confined at scales of order $\Lambda_{H}$ as are the other states, but its mass could well be small, comparable to the $\ell\ell$ pseudoscalar mass. An EFT describing the small mass regime then needs to incorporate the light scalar state together with the pseudoscalars. In the $m_f = 0$ limit, only the pseudoscalar states are massless. The $0^{++}$ decouples at very low energies and $N_f =4$ ChPT should describe the data.

The  dChPT Lagrangian incorporates the effect of a light dilaton state \cite{Golterman:2016lsd,Appelquist:2017vyy,Appelquist:2017wcg,Golterman:2018mfm,Appelquist:2019lgk,Golterman:2020tdq}. While derived for a chirally broken system with degenerate fermions just below the conformal window, we explore its application to our near-conformal mass-split system.

\begin{figure}[tb]
  \centering
  \includegraphics[width=\columnwidth]{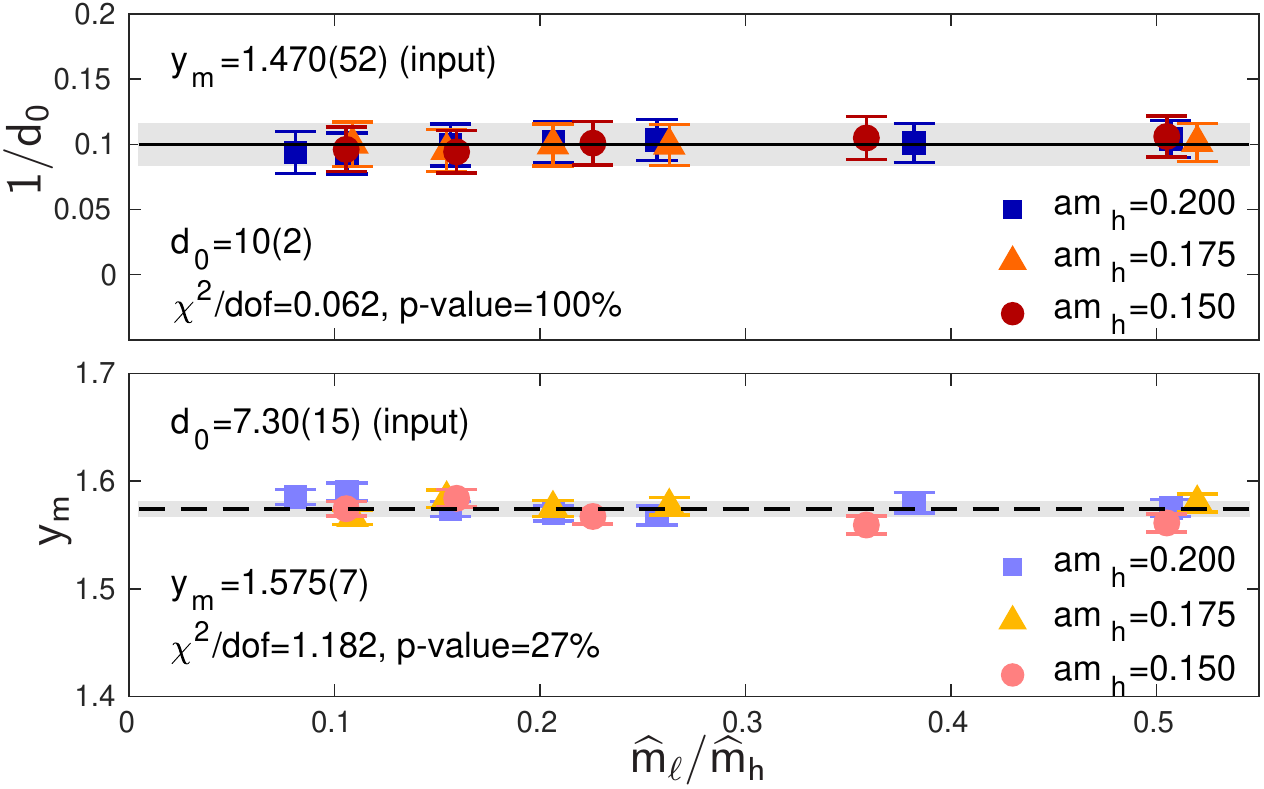}
  \caption{Scaling test of dChPT exploiting Eq.~(\ref{eq:dChPT1lat}). The top panel uses $y_m$ (Eq.~(\ref{eq:ym})) as input to fit $1/d_0$, the bottom panel shows a fit for $y_m$ from a scan over $d_0$ values used as input.}
  \label{Fig.dChPT1}
\end{figure}

dChPT  predicts the scaling relation 
 \begin{align}
d_0 \cdot F_{ps}^{2-y_m} = M_{ps}^2/m_f,
\label{eq:dChPT1} 
\end{align}
 which is a general result first discussed in Refs.~\cite{Golterman:2016lsd,Appelquist:2017wcg} and independent of the specific form of the dilaton effective potential. The quantity $d_0$ is a combination of low energy constants. Using Eq.~(\ref{eq:M_scaling_LH}) we express this relation in terms of lattice quantities of the light sector (dropping the superscripts $\ell\ell$)
\begin{align}
   d_0 \cdot (aF_{ps})^{2-y_m} = (aM_{ps})^2/\widehat m_\ell.
  \label{eq:dChPT1lat} 
\end{align}
From Eq.~(\ref{eq:M_scaling}) we can deduce that $d_0= (aM_{ps})^2\cdot (aF_{ps})^{-2+y_m}/\widehat m_\ell$  may only depend on $\widehat m_\ell/\widehat m_h$, whereas Eq.~(\ref{eq:dChPT1}) states $d_0$ is a constant.

Since our main  goal is to study Eq.~(\ref{eq:dChPT1}),  we simply fix $y_m$ from Eq.~(\ref{eq:ym}) and determine $d_0$ using \eq{eq:dChPT1lat}. As shown in the top panel of Fig.~\ref{Fig.dChPT1}, our data form a flat line without dependence on $\widehat m_\ell/\widehat m_h$.
   A direct fit of our data to Eq.~(\ref{eq:dChPT1lat}) to determine $y_m$ and $d_0$ simultaneously is troublesome because $aF_{ps}$ and $aM_{ps}$  have similar size uncertainties, are highly correlated, and the relation is nonlinear.
Instead we perform a second test scanning a range of input values for $d_0$ and fit for $y_m$. At a minimum $\chi^2/\text{d.o.f.}$ we obtain a $y_m=1.575(7)$ within $2\sigma$ of our reference value and shown in the lower panel of Fig.~\ref{Fig.dChPT1}. In summary, our data are consistent with Eq.~(\ref{eq:dChPT1lat}) and we obtain a rough estimate of $y_m$ and $d_0$.

\begin{figure}[tb]
  \centering
  \includegraphics[width=\columnwidth]{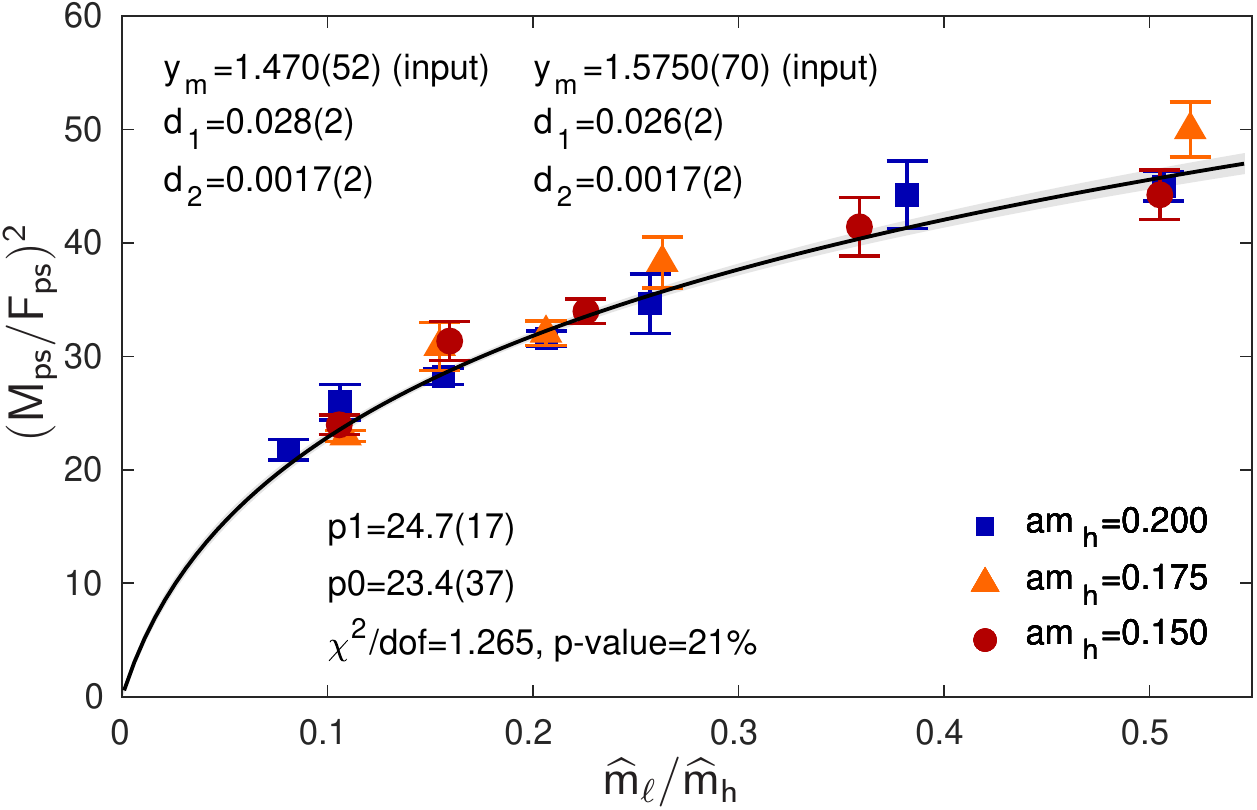}  
  \caption{Test of dChPT using Eq.~(\ref{eq:dChPT2}). The black line with gray band is fitting our ratios $(M_{ps}/F_{ps})^2$ to the function $y= p_0 W_0(p_1\cdot \widehat m_\ell/\widehat m_h)$.}
  \label{Fig.dChPT2}
\end{figure}

Assuming a specific form of the dilaton potential leads to another dChPT relation \cite{Golterman:2020tdq}
 \begin{align}
&\frac{M_{ps}^2}{F_{ps}^2} = \frac{1}{y_m d_1} W_0\left(\frac{y_m d_1}{d_2} m_f\right)
\label{eq:dChPT2}
 \end{align}
 where $W_0$ is the Lambert W-function and $d_1$, $d_2$ are mass independent constants. Figure \ref{Fig.dChPT2} shows a fit of our data to Eq.~(\ref{eq:dChPT2}). The fit has an excellent $p$-value and allows us to determine the constants $d_1$ and  $d_2$. 
 Relations of $N_f=4$ ChPT at leading and next-to-leading order exhibit a mass dependence different from Eqs.~(\ref{eq:dChPT1}) and (\ref{eq:dChPT2}) and do not describe our data.

Finally we comment on the mass dependence of $\sqrt{8t_0}/a$. In ChPT this quantity has a linear mass dependence and  corrections enter only at NNLO \cite{Bar:2013ora}. So far dChPT does not provide a useful description for $\sqrt{8t_0}/a$ \cite{Golterman:2020tdq}. Our results in Fig.~\ref{Fig.hyperscaling_s8t0} show however that $a/\sqrt{8t_0}$ obeys the usual hyperscaling relation in mass-split systems and $a \sqrt{8t_0}\cdot {\widehat m_h}^{-1/y_m}$ is well described by a linear mass dependence.

\section{Conclusion}

In this work we highlight the unique features of the 4+6 mass-split system built on a conformal IRFP. We show that physical masses exhibit hyperscaling and determine the universal mass scaling dimension of the corresponding $N_f=10$ system $y_m=1+\gamma_m^\star=1.47(5)$. 
  This value is smaller than expected for a theory near the edge of the conformal window suggesting that $N_f=9$ or 8 flavor models could be closer to the sill of the conformal window. 

We compare our numerical results to predictions based on dChPT relations and find good agreement. Leading and next-to-leading order standard $N_f=4$ ChPT is, however, not consistent with our data. This strongly suggests that the $0^{++}$ isosinglet scalar of the 4+6 mass-split system is a light state for the investigated parameter range.

There are many important questions to be studied in the future. Numerically determining the $0^{++}$ scalar mass has the highest priority. Investigation of the baryonic anomalous dimension, relevant for partial compositeness, is already in progress \cite{AnomDim:2020}. Calculations of the $S$ parameter and the Higgs potential are planned as well. Finite temperature studies could identify phase transitions with potentially significant implications for the early universe.

\section*{Acknowledgments}\vspace{-2mm}
We are very grateful to Peter Boyle, Guido Cossu, Antonin Portelli, and Azusa Yamaguchi who develop the \texttt{GRID} software library \cite{Gridurl,Boyle:2015tjk} providing the basis of this work and who assisted us in installing and running \texttt{GRID} on different architectures and computing centers. We thank Andrew Pochinsky and Sergey Syritsyn for developing \texttt{QLUA}  \cite{QLUAurl,Pochinsky:2008zz} used for our measurements. The authors thank Maarten Golterman and Yigal Shamir for a critical reading and constructive comments on an early draft of this manuscript. R.C.B.~and C.R.~acknowledge United States Department of Energy (DOE) Award No.~DE-SC0015845. K.C.~acknowledges support from the DOE through the Computational Sciences Graduate Fellowship (DOE CSGF) through grant No.~DE-SC0019323. G.T.F.~acknowledges support from DOE Award No.~DE-SC0019061. A.D.G.~is supported by SNSF grant No.~200021\_17576.  A.H., E.T.N., and O.W.~acknowledge support by DOE Award No.~DE-SC0010005. D.S.~was supported by UK Research and Innovation Future Leader Fellowship No.~{MR/S015418/1}. P.V.~acknowledges the support of the DOE under contract No.~DE-AC52-07NA27344 (LLNL).

We thank the Lawrence Livermore National Laboratory (LLNL) Multiprogrammatic and Institutional Computing program for Grand Challenge supercomputing allocations. We also thank Argonne Leadership Computing Facility for allocations through the INCITE program.  ALCF is supported by DOE contract No.~DE-AC02-06CH11357. Computations for this work were carried out in part on facilities of the USQCD Collaboration, which are funded by the Office of Science of the U.S.~Department of Energy, the RMACC Summit supercomputer \cite{UCsummit}, which is supported by the National Science Foundation (awards No.~ACI-1532235 and No.~ACI-1532236), the University of Colorado Boulder, and Colorado State University and on Boston University computers at the MGHPCC, in part funded by the National Science Foundation (award No.~OCI-1229059).  We thank ANL, BNL, Fermilab, Jefferson Lab, MGHPCC, LLNL, the NSF, the University of Colorado Boulder, and the U.S.~DOE for providing the facilities essential for the completion of this work.

\bibliography{BSM}

\begin{thebibliography}{84}%
\makeatletter
\providecommand \@ifxundefined [1]{%
 \@ifx{#1\undefined}
}%
\providecommand \@ifnum [1]{%
 \ifnum #1\expandafter \@firstoftwo
 \else \expandafter \@secondoftwo
 \fi
}%
\providecommand \@ifx [1]{%
 \ifx #1\expandafter \@firstoftwo
 \else \expandafter \@secondoftwo
 \fi
}%
\providecommand \natexlab [1]{#1}%
\providecommand \enquote  [1]{``#1''}%
\providecommand \bibnamefont  [1]{#1}%
\providecommand \bibfnamefont [1]{#1}%
\providecommand \citenamefont [1]{#1}%
\providecommand \href@noop [0]{\@secondoftwo}%
\providecommand \href [0]{\begingroup \@sanitize@url \@href}%
\providecommand \@href[1]{\@@startlink{#1}\@@href}%
\providecommand \@@href[1]{\endgroup#1\@@endlink}%
\providecommand \@sanitize@url [0]{\catcode `\\12\catcode `\$12\catcode
  `\&12\catcode `\#12\catcode `\^12\catcode `\_12\catcode `\%12\relax}%
\providecommand \@@startlink[1]{}%
\providecommand \@@endlink[0]{}%
\providecommand \url  [0]{\begingroup\@sanitize@url \@url }%
\providecommand \@url [1]{\endgroup\@href {#1}{\urlprefix }}%
\providecommand \urlprefix  [0]{URL }%
\providecommand \Eprint [0]{\href }%
\providecommand \doibase [0]{http://dx.doi.org/}%
\providecommand \selectlanguage [0]{\@gobble}%
\providecommand \bibinfo  [0]{\@secondoftwo}%
\providecommand \bibfield  [0]{\@secondoftwo}%
\providecommand \translation [1]{[#1]}%
\providecommand \BibitemOpen [0]{}%
\providecommand \bibitemStop [0]{}%
\providecommand \bibitemNoStop [0]{.\EOS\space}%
\providecommand \EOS [0]{\spacefactor3000\relax}%
\providecommand \BibitemShut  [1]{\csname bibitem#1\endcsname}%
\let\auto@bib@innerbib\@empty
%</preamble>
\bibitem [{\citenamefont {Aad}\ \emph {et~al.}(2012)\citenamefont {Aad} \emph
  {et~al.}}]{Aad:2012tfa}%
  \BibitemOpen
  \bibfield  {author} {\bibinfo {author} {\bibfnamefont {G.}~\bibnamefont
  {Aad}} \emph {et~al.} (\bibinfo {collaboration} {ATLAS}),\ }\href {\doibase
  10.1016/j.physletb.2012.08.020} {\bibfield  {journal} {\bibinfo  {journal}
  {Phys.Lett.}\ }\textbf {\bibinfo {volume} {B716}},\ \bibinfo {pages} {1}
  (\bibinfo {year} {2012})},\ \Eprint {http://arxiv.org/abs/1207.7214}
  {arXiv:1207.7214 [hep-ex]} \BibitemShut {NoStop}%
%%CITATION = ARXIV:1207.7214;%%
\bibitem [{\citenamefont {Chatrchyan}\ \emph {et~al.}(2012)\citenamefont
  {Chatrchyan} \emph {et~al.}}]{Chatrchyan:2012ufa}%
  \BibitemOpen
  \bibfield  {author} {\bibinfo {author} {\bibfnamefont {S.}~\bibnamefont
  {Chatrchyan}} \emph {et~al.} (\bibinfo {collaboration} {CMS}),\ }\href
  {\doibase 10.1016/j.physletb.2012.08.021} {\bibfield  {journal} {\bibinfo
  {journal} {Phys.Lett.}\ }\textbf {\bibinfo {volume} {B716}},\ \bibinfo
  {pages} {30} (\bibinfo {year} {2012})},\ \Eprint
  {http://arxiv.org/abs/1207.7235} {arXiv:1207.7235 [hep-ex]} \BibitemShut
  {NoStop}%
%%CITATION = ARXIV:1207.7235;%%
\bibitem [{\citenamefont {Aad}\ \emph {et~al.}(2015)\citenamefont {Aad} \emph
  {et~al.}}]{Aad:2015zhl}%
  \BibitemOpen
  \bibfield  {author} {\bibinfo {author} {\bibfnamefont {G.}~\bibnamefont
  {Aad}} \emph {et~al.} (\bibinfo {collaboration} {ATLAS, CMS}),\ }\href
  {\doibase 10.1103/PhysRevLett.114.191803} {\bibfield  {journal} {\bibinfo
  {journal} {Phys. Rev. Lett.}\ }\textbf {\bibinfo {volume} {114}},\ \bibinfo
  {pages} {191803} (\bibinfo {year} {2015})},\ \Eprint
  {http://arxiv.org/abs/1503.07589} {arXiv:1503.07589 [hep-ex]} \BibitemShut
  {NoStop}%
%%CITATION = ARXIV:1503.07589;%%
\bibitem [{\citenamefont {Contino}(2011)}]{Contino:2010rs}%
  \BibitemOpen
  \bibfield  {author} {\bibinfo {author} {\bibfnamefont {R.}~\bibnamefont
  {Contino}},\ }\href {\doibase 10.1142/9789814327183_0005} {\bibfield
  {journal} {\bibinfo  {journal} {{Proceedings TASI}}\ }\textbf {\bibinfo
  {volume} {09}},\ \bibinfo {pages} {235} (\bibinfo {year} {2011})},\ \Eprint
  {http://arxiv.org/abs/1005.4269} {arXiv:1005.4269 [hep-ph]} \BibitemShut
  {NoStop}%
%%CITATION = ARXIV:1005.4269;%%
\bibitem [{\citenamefont {Luty}\ and\ \citenamefont
  {Okui}(2006)}]{Luty:2004ye}%
  \BibitemOpen
  \bibfield  {author} {\bibinfo {author} {\bibfnamefont {M.~A.}\ \bibnamefont
  {Luty}}\ and\ \bibinfo {author} {\bibfnamefont {T.}~\bibnamefont {Okui}},\
  }\href {\doibase 10.1088/1126-6708/2006/09/070} {\bibfield  {journal}
  {\bibinfo  {journal} {JHEP}\ }\textbf {\bibinfo {volume} {09}},\ \bibinfo
  {pages} {070} (\bibinfo {year} {2006})},\ \Eprint
  {http://arxiv.org/abs/hep-ph/0409274} {arXiv:hep-ph/0409274 [hep-ph]}
  \BibitemShut {NoStop}%
%%CITATION = HEP-PH/0409274;%%
\bibitem [{\citenamefont {Dietrich}\ and\ \citenamefont
  {Sannino}(2007)}]{Dietrich:2006cm}%
  \BibitemOpen
  \bibfield  {author} {\bibinfo {author} {\bibfnamefont {D.~D.}\ \bibnamefont
  {Dietrich}}\ and\ \bibinfo {author} {\bibfnamefont {F.}~\bibnamefont
  {Sannino}},\ }\href {\doibase 10.1103/PhysRevD.75.085018} {\bibfield
  {journal} {\bibinfo  {journal} {Phys. Rev.}\ }\textbf {\bibinfo {volume}
  {D75}},\ \bibinfo {pages} {085018} (\bibinfo {year} {2007})},\ \Eprint
  {http://arxiv.org/abs/hep-ph/0611341} {arXiv:hep-ph/0611341 [hep-ph]}
  \BibitemShut {NoStop}%
%%CITATION = HEP-PH/0611341;%%
\bibitem [{\citenamefont {Luty}(2009)}]{Luty:2008vs}%
  \BibitemOpen
  \bibfield  {author} {\bibinfo {author} {\bibfnamefont {M.~A.}\ \bibnamefont
  {Luty}},\ }\href {\doibase 10.1088/1126-6708/2009/04/050} {\bibfield
  {journal} {\bibinfo  {journal} {JHEP}\ }\textbf {\bibinfo {volume} {04}},\
  \bibinfo {pages} {050} (\bibinfo {year} {2009})},\ \Eprint
  {http://arxiv.org/abs/0806.1235} {arXiv:0806.1235 [hep-ph]} \BibitemShut
  {NoStop}%
%%CITATION = ARXIV:0806.1235;%%
\bibitem [{\citenamefont {Brower}\ \emph {et~al.}(2016)\citenamefont {Brower},
  \citenamefont {Hasenfratz}, \citenamefont {Rebbi}, \citenamefont {Weinberg},\
  and\ \citenamefont {Witzel}}]{Brower:2015owo}%
  \BibitemOpen
  \bibfield  {author} {\bibinfo {author} {\bibfnamefont {R.~C.}\ \bibnamefont
  {Brower}}, \bibinfo {author} {\bibfnamefont {A.}~\bibnamefont {Hasenfratz}},
  \bibinfo {author} {\bibfnamefont {C.}~\bibnamefont {Rebbi}}, \bibinfo
  {author} {\bibfnamefont {E.}~\bibnamefont {Weinberg}}, \ and\ \bibinfo
  {author} {\bibfnamefont {O.}~\bibnamefont {Witzel}},\ }\href {\doibase
  10.1103/PhysRevD.93.075028} {\bibfield  {journal} {\bibinfo  {journal} {Phys.
  Rev.}\ }\textbf {\bibinfo {volume} {D93}},\ \bibinfo {pages} {075028}
  (\bibinfo {year} {2016})},\ \Eprint {http://arxiv.org/abs/1512.02576}
  {arXiv:1512.02576 [hep-ph]} \BibitemShut {NoStop}%
%%CITATION = ARXIV:1512.02576;%%
\bibitem [{\citenamefont {Csaki}\ \emph {et~al.}(2016)\citenamefont {Csaki},
  \citenamefont {Grojean},\ and\ \citenamefont {Terning}}]{Csaki:2015hcd}%
  \BibitemOpen
  \bibfield  {author} {\bibinfo {author} {\bibfnamefont {C.}~\bibnamefont
  {Csaki}}, \bibinfo {author} {\bibfnamefont {C.}~\bibnamefont {Grojean}}, \
  and\ \bibinfo {author} {\bibfnamefont {J.}~\bibnamefont {Terning}},\ }\href
  {\doibase 10.1103/RevModPhys.88.045001} {\bibfield  {journal} {\bibinfo
  {journal} {Rev. Mod. Phys.}\ }\textbf {\bibinfo {volume} {88}},\ \bibinfo
  {pages} {045001} (\bibinfo {year} {2016})},\ \Eprint
  {http://arxiv.org/abs/1512.00468} {arXiv:1512.00468 [hep-ph]} \BibitemShut
  {NoStop}%
%%CITATION = ARXIV:1512.00468;%%
\bibitem [{\citenamefont {Arkani-Hamed}\ \emph {et~al.}(2016)\citenamefont
  {Arkani-Hamed}, \citenamefont {D'Agnolo}, \citenamefont {Low},\ and\
  \citenamefont {Pinner}}]{Arkani-Hamed:2016kpz}%
  \BibitemOpen
  \bibfield  {author} {\bibinfo {author} {\bibfnamefont {N.}~\bibnamefont
  {Arkani-Hamed}}, \bibinfo {author} {\bibfnamefont {R.~T.}\ \bibnamefont
  {D'Agnolo}}, \bibinfo {author} {\bibfnamefont {M.}~\bibnamefont {Low}}, \
  and\ \bibinfo {author} {\bibfnamefont {D.}~\bibnamefont {Pinner}},\ }\href
  {\doibase 10.1007/JHEP11(2016)082} {\bibfield  {journal} {\bibinfo  {journal}
  {JHEP}\ }\textbf {\bibinfo {volume} {11}},\ \bibinfo {pages} {082} (\bibinfo
  {year} {2016})},\ \Eprint {http://arxiv.org/abs/1608.01675} {arXiv:1608.01675
  [hep-ph]} \BibitemShut {NoStop}%
%%CITATION = ARXIV:1608.01675;%%
\bibitem [{\citenamefont {Witzel}(2019)}]{Witzel:2019jbe}%
  \BibitemOpen
  \bibfield  {author} {\bibinfo {author} {\bibfnamefont {O.}~\bibnamefont
  {Witzel}},\ }\href {\doibase 10.22323/1.334.0006} {\bibfield  {journal}
  {\bibinfo  {journal} {PoS}\ }\textbf {\bibinfo {volume} {LATTICE2018}},\
  \bibinfo {pages} {006} (\bibinfo {year} {2019})},\ \Eprint
  {http://arxiv.org/abs/1901.08216} {arXiv:1901.08216 [hep-lat]} \BibitemShut
  {NoStop}%
%%CITATION = ARXIV:1901.08216;%%
\bibitem [{\citenamefont {Yamawaki}\ \emph {et~al.}(1986)\citenamefont
  {Yamawaki}, \citenamefont {Bando},\ and\ \citenamefont
  {Matumoto}}]{Yamawaki:1985zg}%
  \BibitemOpen
  \bibfield  {author} {\bibinfo {author} {\bibfnamefont {K.}~\bibnamefont
  {Yamawaki}}, \bibinfo {author} {\bibfnamefont {M.}~\bibnamefont {Bando}}, \
  and\ \bibinfo {author} {\bibfnamefont {K.-i.}\ \bibnamefont {Matumoto}},\
  }\href {\doibase 10.1103/PhysRevLett.56.1335} {\bibfield  {journal} {\bibinfo
   {journal} {Phys. Rev. Lett.}\ }\textbf {\bibinfo {volume} {56}},\ \bibinfo
  {pages} {1335} (\bibinfo {year} {1986})}\BibitemShut {NoStop}%
%%CITATION = PRLTA,56,1335;%%
\bibitem [{\citenamefont {Appelquist}\ \emph {et~al.}(1986)\citenamefont
  {Appelquist}, \citenamefont {Karabali},\ and\ \citenamefont
  {Wijewardhana}}]{Appelquist:1986an}%
  \BibitemOpen
  \bibfield  {author} {\bibinfo {author} {\bibfnamefont {T.~W.}\ \bibnamefont
  {Appelquist}}, \bibinfo {author} {\bibfnamefont {D.}~\bibnamefont
  {Karabali}}, \ and\ \bibinfo {author} {\bibfnamefont {L.}~\bibnamefont
  {Wijewardhana}},\ }\href {\doibase 10.1103/PhysRevLett.57.957} {\bibfield
  {journal} {\bibinfo  {journal} {Phys. Rev. Lett.}\ }\textbf {\bibinfo
  {volume} {57}},\ \bibinfo {pages} {957} (\bibinfo {year} {1986})}\BibitemShut
  {NoStop}%
\bibitem [{\citenamefont {Bando}\ \emph {et~al.}(1988)\citenamefont {Bando},
  \citenamefont {Kugo},\ and\ \citenamefont {Yamawaki}}]{Bando:1987br}%
  \BibitemOpen
  \bibfield  {author} {\bibinfo {author} {\bibfnamefont {M.}~\bibnamefont
  {Bando}}, \bibinfo {author} {\bibfnamefont {T.}~\bibnamefont {Kugo}}, \ and\
  \bibinfo {author} {\bibfnamefont {K.}~\bibnamefont {Yamawaki}},\ }\href
  {\doibase 10.1016/0370-1573(88)90019-1} {\bibfield  {journal} {\bibinfo
  {journal} {Phys. Rept.}\ }\textbf {\bibinfo {volume} {164}},\ \bibinfo
  {pages} {217} (\bibinfo {year} {1988})}\BibitemShut {NoStop}%
%%CITATION = PRPLC,164,217;%%
\bibitem [{\citenamefont {Brower}\ \emph {et~al.}(2015)\citenamefont {Brower},
  \citenamefont {Hasenfratz}, \citenamefont {Rebbi}, \citenamefont {Weinberg},\
  and\ \citenamefont {Witzel}}]{Brower:2014dfa}%
  \BibitemOpen
  \bibfield  {author} {\bibinfo {author} {\bibfnamefont {R.~C.}\ \bibnamefont
  {Brower}}, \bibinfo {author} {\bibfnamefont {A.}~\bibnamefont {Hasenfratz}},
  \bibinfo {author} {\bibfnamefont {C.}~\bibnamefont {Rebbi}}, \bibinfo
  {author} {\bibfnamefont {E.}~\bibnamefont {Weinberg}}, \ and\ \bibinfo
  {author} {\bibfnamefont {O.}~\bibnamefont {Witzel}},\ }\href {\doibase
  10.1134/S1063776115030176} {\bibfield  {journal} {\bibinfo  {journal} {J.
  Exp. Theor. Phys.}\ }\textbf {\bibinfo {volume} {120}},\ \bibinfo {pages}
  {423} (\bibinfo {year} {2015})},\ \Eprint {http://arxiv.org/abs/1410.4091}
  {arXiv:1410.4091 [hep-lat]} \BibitemShut {NoStop}%
%%CITATION = ARXIV:1410.4091;%%
\bibitem [{\citenamefont {Hasenfratz}\ \emph
  {et~al.}(2017{\natexlab{a}})\citenamefont {Hasenfratz}, \citenamefont
  {Rebbi},\ and\ \citenamefont {Witzel}}]{Hasenfratz:2016gut}%
  \BibitemOpen
  \bibfield  {author} {\bibinfo {author} {\bibfnamefont {A.}~\bibnamefont
  {Hasenfratz}}, \bibinfo {author} {\bibfnamefont {C.}~\bibnamefont {Rebbi}}, \
  and\ \bibinfo {author} {\bibfnamefont {O.}~\bibnamefont {Witzel}},\ }\href
  {\doibase 10.1016/j.physletb.2017.07.058} {\bibfield  {journal} {\bibinfo
  {journal} {Phys. Lett.}\ }\textbf {\bibinfo {volume} {B773}},\ \bibinfo
  {pages} {86} (\bibinfo {year} {2017}{\natexlab{a}})},\ \Eprint
  {http://arxiv.org/abs/1609.01401} {arXiv:1609.01401 [hep-ph]} \BibitemShut
  {NoStop}%
%%CITATION = ARXIV:1609.01401;%%
\bibitem [{\citenamefont {Hasenfratz}\ \emph {et~al.}(2018)\citenamefont
  {Hasenfratz}, \citenamefont {Rebbi},\ and\ \citenamefont
  {Witzel}}]{Hasenfratz:2017hdd}%
  \BibitemOpen
  \bibfield  {author} {\bibinfo {author} {\bibfnamefont {A.}~\bibnamefont
  {Hasenfratz}}, \bibinfo {author} {\bibfnamefont {C.}~\bibnamefont {Rebbi}}, \
  and\ \bibinfo {author} {\bibfnamefont {O.}~\bibnamefont {Witzel}},\ }\href
  {\doibase 10.1051/epjconf/201817508007} {\bibfield  {journal} {\bibinfo
  {journal} {EPJ Web Conf.}\ }\textbf {\bibinfo {volume} {175}},\ \bibinfo
  {pages} {08007} (\bibinfo {year} {2018})},\ \Eprint
  {http://arxiv.org/abs/1710.08970} {arXiv:1710.08970 [hep-lat]} \BibitemShut
  {NoStop}%
\bibitem [{\citenamefont {Miransky}(1999)}]{Miransky:1998dh}%
  \BibitemOpen
  \bibfield  {author} {\bibinfo {author} {\bibfnamefont {V.~A.}\ \bibnamefont
  {Miransky}},\ }\href {\doibase 10.1103/PhysRevD.59.105003} {\bibfield
  {journal} {\bibinfo  {journal} {Phys. Rev.}\ }\textbf {\bibinfo {volume}
  {D59}},\ \bibinfo {pages} {105003} (\bibinfo {year} {1999})},\ \Eprint
  {http://arxiv.org/abs/hep-ph/9812350} {arXiv:hep-ph/9812350 [hep-ph]}
  \BibitemShut {NoStop}%
%%CITATION = HEP-PH/9812350;%%
\bibitem [{\citenamefont {Aoki}\ \emph {et~al.}(2013)\citenamefont {Aoki},
  \citenamefont {Aoyama}, \citenamefont {Kurachi}, \citenamefont {Maskawa},
  \citenamefont {Nagai}, \citenamefont {Ohki}, \citenamefont {Rinaldi},
  \citenamefont {Shibata}, \citenamefont {Yamawaki},\ and\ \citenamefont
  {Yamazaki}}]{Aoki:2013zsa}%
  \BibitemOpen
  \bibfield  {author} {\bibinfo {author} {\bibfnamefont {Y.}~\bibnamefont
  {Aoki}}, \bibinfo {author} {\bibfnamefont {T.}~\bibnamefont {Aoyama}},
  \bibinfo {author} {\bibfnamefont {M.}~\bibnamefont {Kurachi}}, \bibinfo
  {author} {\bibfnamefont {T.}~\bibnamefont {Maskawa}}, \bibinfo {author}
  {\bibfnamefont {K.-i.}\ \bibnamefont {Nagai}}, \bibinfo {author}
  {\bibfnamefont {H.}~\bibnamefont {Ohki}}, \bibinfo {author} {\bibfnamefont
  {E.}~\bibnamefont {Rinaldi}}, \bibinfo {author} {\bibfnamefont
  {A.}~\bibnamefont {Shibata}}, \bibinfo {author} {\bibfnamefont
  {K.}~\bibnamefont {Yamawaki}}, \ and\ \bibinfo {author} {\bibfnamefont
  {T.}~\bibnamefont {Yamazaki}} (\bibinfo {collaboration} {LatKMI}),\ }\href
  {\doibase 10.1103/PhysRevLett.111.162001} {\bibfield  {journal} {\bibinfo
  {journal} {Phys. Rev. Lett.}\ }\textbf {\bibinfo {volume} {111}},\ \bibinfo
  {pages} {162001} (\bibinfo {year} {2013})},\ \Eprint
  {http://arxiv.org/abs/1305.6006} {arXiv:1305.6006 [hep-lat]} \BibitemShut
  {NoStop}%
%%CITATION = ARXIV:1305.6006;%%
\bibitem [{\citenamefont {Golterman}\ and\ \citenamefont
  {Shamir}(2016)}]{Golterman:2016lsd}%
  \BibitemOpen
  \bibfield  {author} {\bibinfo {author} {\bibfnamefont {M.}~\bibnamefont
  {Golterman}}\ and\ \bibinfo {author} {\bibfnamefont {Y.}~\bibnamefont
  {Shamir}},\ }\href {\doibase 10.1103/PhysRevD.94.054502} {\bibfield
  {journal} {\bibinfo  {journal} {Phys. Rev.}\ }\textbf {\bibinfo {volume}
  {D94}},\ \bibinfo {pages} {054502} (\bibinfo {year} {2016})},\ \Eprint
  {http://arxiv.org/abs/1603.04575} {arXiv:1603.04575 [hep-ph]} \BibitemShut
  {NoStop}%
%%CITATION = ARXIV:1603.04575;%%
\bibitem [{\citenamefont {Appelquist}\ \emph {et~al.}(2018)\citenamefont
  {Appelquist}, \citenamefont {Ingoldby},\ and\ \citenamefont
  {Piai}}]{Appelquist:2017vyy}%
  \BibitemOpen
  \bibfield  {author} {\bibinfo {author} {\bibfnamefont {T.}~\bibnamefont
  {Appelquist}}, \bibinfo {author} {\bibfnamefont {J.}~\bibnamefont
  {Ingoldby}}, \ and\ \bibinfo {author} {\bibfnamefont {M.}~\bibnamefont
  {Piai}},\ }\href {\doibase 10.1007/JHEP03(2018)039} {\bibfield  {journal}
  {\bibinfo  {journal} {JHEP}\ }\textbf {\bibinfo {volume} {03}},\ \bibinfo
  {pages} {039} (\bibinfo {year} {2018})},\ \Eprint
  {http://arxiv.org/abs/1711.00067} {arXiv:1711.00067 [hep-ph]} \BibitemShut
  {NoStop}%
%%CITATION = ARXIV:1711.00067;%%
\bibitem [{\citenamefont {Appelquist}\ \emph {et~al.}(2017)\citenamefont
  {Appelquist}, \citenamefont {Ingoldby},\ and\ \citenamefont
  {Piai}}]{Appelquist:2017wcg}%
  \BibitemOpen
  \bibfield  {author} {\bibinfo {author} {\bibfnamefont {T.}~\bibnamefont
  {Appelquist}}, \bibinfo {author} {\bibfnamefont {J.}~\bibnamefont
  {Ingoldby}}, \ and\ \bibinfo {author} {\bibfnamefont {M.}~\bibnamefont
  {Piai}},\ }\href {\doibase 10.1007/JHEP07(2017)035} {\bibfield  {journal}
  {\bibinfo  {journal} {JHEP}\ }\textbf {\bibinfo {volume} {07}},\ \bibinfo
  {pages} {035} (\bibinfo {year} {2017})},\ \Eprint
  {http://arxiv.org/abs/1702.04410} {arXiv:1702.04410 [hep-ph]} \BibitemShut
  {NoStop}%
%%CITATION = ARXIV:1702.04410;%%
\bibitem [{\citenamefont {Golterman}\ and\ \citenamefont
  {Shamir}(2018)}]{Golterman:2018mfm}%
  \BibitemOpen
  \bibfield  {author} {\bibinfo {author} {\bibfnamefont {M.}~\bibnamefont
  {Golterman}}\ and\ \bibinfo {author} {\bibfnamefont {Y.}~\bibnamefont
  {Shamir}},\ }\href {\doibase 10.1103/PhysRevD.98.056025} {\bibfield
  {journal} {\bibinfo  {journal} {Phys. Rev.}\ }\textbf {\bibinfo {volume}
  {D98}},\ \bibinfo {pages} {056025} (\bibinfo {year} {2018})},\ \Eprint
  {http://arxiv.org/abs/1805.00198} {arXiv:1805.00198 [hep-ph]} \BibitemShut
  {NoStop}%
%%CITATION = ARXIV:1805.00198;%%
\bibitem [{\citenamefont {Appelquist}\ \emph {et~al.}(2020)\citenamefont
  {Appelquist}, \citenamefont {Ingoldby},\ and\ \citenamefont
  {Piai}}]{Appelquist:2019lgk}%
  \BibitemOpen
  \bibfield  {author} {\bibinfo {author} {\bibfnamefont {T.}~\bibnamefont
  {Appelquist}}, \bibinfo {author} {\bibfnamefont {J.}~\bibnamefont
  {Ingoldby}}, \ and\ \bibinfo {author} {\bibfnamefont {M.}~\bibnamefont
  {Piai}},\ }\href {\doibase 10.1103/PhysRevD.101.075025} {\bibfield  {journal}
  {\bibinfo  {journal} {Phys. Rev. D}\ }\textbf {\bibinfo {volume} {101}},\
  \bibinfo {pages} {075025} (\bibinfo {year} {2020})},\ \Eprint
  {http://arxiv.org/abs/1908.00895} {arXiv:1908.00895 [hep-ph]} \BibitemShut
  {NoStop}%
\bibitem [{\citenamefont {Golterman}\ \emph {et~al.}(2020)\citenamefont
  {Golterman}, \citenamefont {Neil},\ and\ \citenamefont
  {Shamir}}]{Golterman:2020tdq}%
  \BibitemOpen
  \bibfield  {author} {\bibinfo {author} {\bibfnamefont {M.}~\bibnamefont
  {Golterman}}, \bibinfo {author} {\bibfnamefont {E.~T.}\ \bibnamefont {Neil}},
  \ and\ \bibinfo {author} {\bibfnamefont {Y.}~\bibnamefont {Shamir}},\ }\href
  {\doibase 10.1103/PhysRevD.102.034515} {\bibfield  {journal} {\bibinfo
  {journal} {Phys. Rev. D}\ }\textbf {\bibinfo {volume} {102}},\ \bibinfo
  {pages} {034515} (\bibinfo {year} {2020})},\ \Eprint
  {http://arxiv.org/abs/2003.00114} {arXiv:2003.00114 [hep-ph]} \BibitemShut
  {NoStop}%
\bibitem [{\citenamefont {Hayakawa}\ \emph {et~al.}(2011)\citenamefont
  {Hayakawa}, \citenamefont {Ishikawa}, \citenamefont {Osaki}, \citenamefont
  {Takeda}, \citenamefont {Uno},\ and\ \citenamefont
  {Yamada}}]{Hayakawa:2010yn}%
  \BibitemOpen
  \bibfield  {author} {\bibinfo {author} {\bibfnamefont {M.}~\bibnamefont
  {Hayakawa}}, \bibinfo {author} {\bibfnamefont {K.-I.}\ \bibnamefont
  {Ishikawa}}, \bibinfo {author} {\bibfnamefont {Y.}~\bibnamefont {Osaki}},
  \bibinfo {author} {\bibfnamefont {S.}~\bibnamefont {Takeda}}, \bibinfo
  {author} {\bibfnamefont {S.}~\bibnamefont {Uno}}, \ and\ \bibinfo {author}
  {\bibfnamefont {N.}~\bibnamefont {Yamada}},\ }\href {\doibase
  10.1103/PhysRevD.83.074509} {\bibfield  {journal} {\bibinfo  {journal} {Phys.
  Rev. D}\ }\textbf {\bibinfo {volume} {83}},\ \bibinfo {pages} {074509}
  (\bibinfo {year} {2011})},\ \Eprint {http://arxiv.org/abs/1011.2577}
  {arXiv:1011.2577 [hep-lat]} \BibitemShut {NoStop}%
\bibitem [{\citenamefont {Appelquist}\ \emph {et~al.}(2012)\citenamefont
  {Appelquist} \emph {et~al.}}]{Appelquist:2012nz}%
  \BibitemOpen
  \bibfield  {author} {\bibinfo {author} {\bibfnamefont {T.}~\bibnamefont
  {Appelquist}} \emph {et~al.},\ }\href@noop {} {\  (\bibinfo {year} {2012})},\
  \Eprint {http://arxiv.org/abs/1204.6000} {arXiv:1204.6000 [hep-ph]}
  \BibitemShut {NoStop}%
\bibitem [{\citenamefont {Chiu}(2016)}]{Chiu:2016uui}%
  \BibitemOpen
  \bibfield  {author} {\bibinfo {author} {\bibfnamefont {T.-W.}\ \bibnamefont
  {Chiu}},\ }\href@noop {} {\  (\bibinfo {year} {2016})},\ \Eprint
  {http://arxiv.org/abs/1603.08854} {arXiv:1603.08854 [hep-lat]} \BibitemShut
  {NoStop}%
%%CITATION = ARXIV:1603.08854;%%
\bibitem [{\citenamefont {Chiu}(2017)}]{Chiu:2017kza}%
  \BibitemOpen
  \bibfield  {author} {\bibinfo {author} {\bibfnamefont {T.-W.}\ \bibnamefont
  {Chiu}},\ }\href {\doibase 10.22323/1.256.0228} {\bibfield  {journal}
  {\bibinfo  {journal} {PoS}\ }\textbf {\bibinfo {volume} {LATTICE2016}},\
  \bibinfo {pages} {228} (\bibinfo {year} {2017})}\BibitemShut {NoStop}%
%%CITATION = POSCI,LATTICE2016,228;%%
\bibitem [{\citenamefont {Chiu}(2019)}]{Chiu:2018edw}%
  \BibitemOpen
  \bibfield  {author} {\bibinfo {author} {\bibfnamefont {T.-W.}\ \bibnamefont
  {Chiu}},\ }\href@noop {} {\bibfield  {journal} {\bibinfo  {journal} {Phys.
  Rev.}\ }\textbf {\bibinfo {volume} {D99}},\ \bibinfo {pages} {014507}
  (\bibinfo {year} {2019})},\ \Eprint {http://arxiv.org/abs/1811.01729}
  {arXiv:1811.01729 [hep-lat]} \BibitemShut {NoStop}%
%%CITATION = ARXIV:1811.01729;%%
\bibitem [{\citenamefont {Fodor}\ \emph {et~al.}(2018)\citenamefont {Fodor},
  \citenamefont {Holland}, \citenamefont {Kuti}, \citenamefont {Nogradi},\ and\
  \citenamefont {Wong}}]{Fodor:2017gtj}%
  \BibitemOpen
  \bibfield  {author} {\bibinfo {author} {\bibfnamefont {Z.}~\bibnamefont
  {Fodor}}, \bibinfo {author} {\bibfnamefont {K.}~\bibnamefont {Holland}},
  \bibinfo {author} {\bibfnamefont {J.}~\bibnamefont {Kuti}}, \bibinfo {author}
  {\bibfnamefont {D.}~\bibnamefont {Nogradi}}, \ and\ \bibinfo {author}
  {\bibfnamefont {C.~H.}\ \bibnamefont {Wong}},\ }\href {\doibase
  10.1016/j.physletb.2018.02.008} {\bibfield  {journal} {\bibinfo  {journal}
  {Phys. Lett.}\ }\textbf {\bibinfo {volume} {B779}},\ \bibinfo {pages} {230}
  (\bibinfo {year} {2018})},\ \Eprint {http://arxiv.org/abs/1710.09262}
  {arXiv:1710.09262 [hep-lat]} \BibitemShut {NoStop}%
%%CITATION = ARXIV:1710.09262;%%
\bibitem [{\citenamefont {Hasenfratz}\ \emph {et~al.}(2019)\citenamefont
  {Hasenfratz}, \citenamefont {Rebbi},\ and\ \citenamefont
  {Witzel}}]{Hasenfratz:2017qyr}%
  \BibitemOpen
  \bibfield  {author} {\bibinfo {author} {\bibfnamefont {A.}~\bibnamefont
  {Hasenfratz}}, \bibinfo {author} {\bibfnamefont {C.}~\bibnamefont {Rebbi}}, \
  and\ \bibinfo {author} {\bibfnamefont {O.}~\bibnamefont {Witzel}},\ }\href
  {\doibase 10.1016/j.physletb.2019.134937} {\bibfield  {journal} {\bibinfo
  {journal} {Phys. Lett.}\ }\textbf {\bibinfo {volume} {B798}},\ \bibinfo
  {pages} {134937} (\bibinfo {year} {2019})},\ \Eprint
  {http://arxiv.org/abs/1710.11578} {arXiv:1710.11578 [hep-lat]} \BibitemShut
  {NoStop}%
%%CITATION = ARXIV:1710.11578;%%
\bibitem [{\citenamefont {Hasenfratz}\ \emph {et~al.}(2020)\citenamefont
  {Hasenfratz}, \citenamefont {Rebbi},\ and\ \citenamefont
  {Witzel}}]{Hasenfratz:2020ess}%
  \BibitemOpen
  \bibfield  {author} {\bibinfo {author} {\bibfnamefont {A.}~\bibnamefont
  {Hasenfratz}}, \bibinfo {author} {\bibfnamefont {C.}~\bibnamefont {Rebbi}}, \
  and\ \bibinfo {author} {\bibfnamefont {O.}~\bibnamefont {Witzel}},\ }\href
  {\doibase 10.1103/PhysRevD.101.114508} {\bibfield  {journal} {\bibinfo
  {journal} {Phys. Rev. D}\ }\textbf {\bibinfo {volume} {101}},\ \bibinfo
  {pages} {114508} (\bibinfo {year} {2020})},\ \Eprint
  {http://arxiv.org/abs/2004.00754} {arXiv:2004.00754 [hep-lat]} \BibitemShut
  {NoStop}%
\bibitem [{\citenamefont {Baikov}\ \emph {et~al.}(2017)\citenamefont {Baikov},
  \citenamefont {Chetyrkin},\ and\ \citenamefont {Kühn}}]{Baikov:2016tgj}%
  \BibitemOpen
  \bibfield  {author} {\bibinfo {author} {\bibfnamefont {P.~A.}\ \bibnamefont
  {Baikov}}, \bibinfo {author} {\bibfnamefont {K.~G.}\ \bibnamefont
  {Chetyrkin}}, \ and\ \bibinfo {author} {\bibfnamefont {J.~H.}\ \bibnamefont
  {Kühn}},\ }\href {\doibase 10.1103/PhysRevLett.118.082002} {\bibfield
  {journal} {\bibinfo  {journal} {Phys. Rev. Lett.}\ }\textbf {\bibinfo
  {volume} {118}},\ \bibinfo {pages} {082002} (\bibinfo {year} {2017})},\
  \Eprint {http://arxiv.org/abs/1606.08659} {arXiv:1606.08659 [hep-ph]}
  \BibitemShut {NoStop}%
%%CITATION = ARXIV:1606.08659;%%
\bibitem [{\citenamefont {Ryttov}\ and\ \citenamefont
  {Shrock}(2011)}]{Ryttov:2010iz}%
  \BibitemOpen
  \bibfield  {author} {\bibinfo {author} {\bibfnamefont {T.~A.}\ \bibnamefont
  {Ryttov}}\ and\ \bibinfo {author} {\bibfnamefont {R.}~\bibnamefont
  {Shrock}},\ }\href {\doibase 10.1103/PhysRevD.83.056011} {\bibfield
  {journal} {\bibinfo  {journal} {Phys. Rev.}\ }\textbf {\bibinfo {volume}
  {D83}},\ \bibinfo {pages} {056011} (\bibinfo {year} {2011})},\ \Eprint
  {http://arxiv.org/abs/1011.4542} {arXiv:1011.4542} \BibitemShut {NoStop}%
\bibitem [{\citenamefont {Ryttov}\ and\ \citenamefont
  {Shrock}(2016{\natexlab{a}})}]{Ryttov:2016ner}%
  \BibitemOpen
  \bibfield  {author} {\bibinfo {author} {\bibfnamefont {T.~A.}\ \bibnamefont
  {Ryttov}}\ and\ \bibinfo {author} {\bibfnamefont {R.}~\bibnamefont
  {Shrock}},\ }\href {\doibase 10.1103/PhysRevD.94.105015} {\bibfield
  {journal} {\bibinfo  {journal} {Phys. Rev.}\ }\textbf {\bibinfo {volume}
  {D94}},\ \bibinfo {pages} {105015} (\bibinfo {year} {2016}{\natexlab{a}})},\
  \Eprint {http://arxiv.org/abs/1607.06866} {arXiv:1607.06866 [hep-th]}
  \BibitemShut {NoStop}%
%%CITATION = ARXIV:1607.06866;%%
\bibitem [{\citenamefont {Ryttov}\ and\ \citenamefont
  {Shrock}(2016{\natexlab{b}})}]{Ryttov:2016hal}%
  \BibitemOpen
  \bibfield  {author} {\bibinfo {author} {\bibfnamefont {T.~A.}\ \bibnamefont
  {Ryttov}}\ and\ \bibinfo {author} {\bibfnamefont {R.}~\bibnamefont
  {Shrock}},\ }\href {\doibase 10.1103/PhysRevD.94.125005} {\bibfield
  {journal} {\bibinfo  {journal} {Phys. Rev.}\ }\textbf {\bibinfo {volume}
  {D94}},\ \bibinfo {pages} {125005} (\bibinfo {year} {2016}{\natexlab{b}})},\
  \Eprint {http://arxiv.org/abs/1610.00387} {arXiv:1610.00387 [hep-th]}
  \BibitemShut {NoStop}%
%%CITATION = ARXIV:1610.00387;%%
\bibitem [{\citenamefont {Ryttov}\ and\ \citenamefont
  {Shrock}(2017)}]{Ryttov:2017kmx}%
  \BibitemOpen
  \bibfield  {author} {\bibinfo {author} {\bibfnamefont {T.~A.}\ \bibnamefont
  {Ryttov}}\ and\ \bibinfo {author} {\bibfnamefont {R.}~\bibnamefont
  {Shrock}},\ }\href {\doibase 10.1103/PhysRevD.95.105004} {\bibfield
  {journal} {\bibinfo  {journal} {Phys. Rev.}\ }\textbf {\bibinfo {volume}
  {D95}},\ \bibinfo {pages} {105004} (\bibinfo {year} {2017})},\ \Eprint
  {http://arxiv.org/abs/1703.08558} {arXiv:1703.08558 [hep-th]} \BibitemShut
  {NoStop}%
%%CITATION = ARXIV:1703.08558;%%
\bibitem [{\citenamefont {Antipin}\ \emph {et~al.}(2019)\citenamefont
  {Antipin}, \citenamefont {Maiezza},\ and\ \citenamefont
  {Vasquez}}]{Antipin:2018asc}%
  \BibitemOpen
  \bibfield  {author} {\bibinfo {author} {\bibfnamefont {O.}~\bibnamefont
  {Antipin}}, \bibinfo {author} {\bibfnamefont {A.}~\bibnamefont {Maiezza}}, \
  and\ \bibinfo {author} {\bibfnamefont {J.~C.}\ \bibnamefont {Vasquez}},\
  }\href {\doibase 10.1016/j.nuclphysb.2019.02.014} {\bibfield  {journal}
  {\bibinfo  {journal} {Nucl. Phys. B}\ }\textbf {\bibinfo {volume} {941}},\
  \bibinfo {pages} {72} (\bibinfo {year} {2019})},\ \Eprint
  {http://arxiv.org/abs/1807.05060} {arXiv:1807.05060 [hep-th]} \BibitemShut
  {NoStop}%
\bibitem [{\citenamefont {Di~Pietro}\ and\ \citenamefont
  {Serone}(2020)}]{DiPietro:2020jne}%
  \BibitemOpen
  \bibfield  {author} {\bibinfo {author} {\bibfnamefont {L.}~\bibnamefont
  {Di~Pietro}}\ and\ \bibinfo {author} {\bibfnamefont {M.}~\bibnamefont
  {Serone}},\ }\href {\doibase 10.1007/JHEP07(2020)049} {\bibfield  {journal}
  {\bibinfo  {journal} {JHEP}\ }\textbf {\bibinfo {volume} {07}},\ \bibinfo
  {pages} {049} (\bibinfo {year} {2020})},\ \Eprint
  {http://arxiv.org/abs/2003.01742} {arXiv:2003.01742 [hep-th]} \BibitemShut
  {NoStop}%
\bibitem [{\citenamefont {Ma}\ and\ \citenamefont
  {Cacciapaglia}(2016)}]{Ma:2015gra}%
  \BibitemOpen
  \bibfield  {author} {\bibinfo {author} {\bibfnamefont {T.}~\bibnamefont
  {Ma}}\ and\ \bibinfo {author} {\bibfnamefont {G.}~\bibnamefont
  {Cacciapaglia}},\ }\href {\doibase 10.1007/JHEP03(2016)211} {\bibfield
  {journal} {\bibinfo  {journal} {JHEP}\ }\textbf {\bibinfo {volume} {03}},\
  \bibinfo {pages} {211} (\bibinfo {year} {2016})},\ \Eprint
  {http://arxiv.org/abs/1508.07014} {arXiv:1508.07014 [hep-ph]} \BibitemShut
  {NoStop}%
%%CITATION = ARXIV:1508.07014;%%
\bibitem [{\citenamefont {Buarque~Franzosi}\ \emph {et~al.}(2020)\citenamefont
  {Buarque~Franzosi}, \citenamefont {Cacciapaglia},\ and\ \citenamefont
  {Deandrea}}]{BuarqueFranzosi:2018eaj}%
  \BibitemOpen
  \bibfield  {author} {\bibinfo {author} {\bibfnamefont {D.}~\bibnamefont
  {Buarque~Franzosi}}, \bibinfo {author} {\bibfnamefont {G.}~\bibnamefont
  {Cacciapaglia}}, \ and\ \bibinfo {author} {\bibfnamefont {A.}~\bibnamefont
  {Deandrea}},\ }\href {\doibase 10.1140/epjc/s10052-019-7572-z} {\bibfield
  {journal} {\bibinfo  {journal} {Eur. Phys. J. C}\ }\textbf {\bibinfo {volume}
  {80}},\ \bibinfo {pages} {28} (\bibinfo {year} {2020})},\ \Eprint
  {http://arxiv.org/abs/1809.09146} {arXiv:1809.09146 [hep-ph]} \BibitemShut
  {NoStop}%
\bibitem [{\citenamefont {Marzocca}(2018)}]{Marzocca:2018wcf}%
  \BibitemOpen
  \bibfield  {author} {\bibinfo {author} {\bibfnamefont {D.}~\bibnamefont
  {Marzocca}},\ }\href {\doibase 10.1007/JHEP07(2018)121} {\bibfield  {journal}
  {\bibinfo  {journal} {JHEP}\ }\textbf {\bibinfo {volume} {07}},\ \bibinfo
  {pages} {121} (\bibinfo {year} {2018})},\ \Eprint
  {http://arxiv.org/abs/1803.10972} {arXiv:1803.10972 [hep-ph]} \BibitemShut
  {NoStop}%
%%CITATION = ARXIV:1803.10972;%%
\bibitem [{\citenamefont {Vecchi}(2017)}]{Vecchi:2015fma}%
  \BibitemOpen
  \bibfield  {author} {\bibinfo {author} {\bibfnamefont {L.}~\bibnamefont
  {Vecchi}},\ }\href {\doibase 10.1007/JHEP02(2017)094} {\bibfield  {journal}
  {\bibinfo  {journal} {JHEP}\ }\textbf {\bibinfo {volume} {02}},\ \bibinfo
  {pages} {094} (\bibinfo {year} {2017})},\ \Eprint
  {http://arxiv.org/abs/1506.00623} {arXiv:1506.00623 [hep-ph]} \BibitemShut
  {NoStop}%
%%CITATION = ARXIV:1506.00623;%%
\bibitem [{\citenamefont {Ferretti}\ and\ \citenamefont
  {Karateev}(2014)}]{Ferretti:2013kya}%
  \BibitemOpen
  \bibfield  {author} {\bibinfo {author} {\bibfnamefont {G.}~\bibnamefont
  {Ferretti}}\ and\ \bibinfo {author} {\bibfnamefont {D.}~\bibnamefont
  {Karateev}},\ }\href {\doibase 10.1007/JHEP03(2014)077} {\bibfield  {journal}
  {\bibinfo  {journal} {JHEP}\ }\textbf {\bibinfo {volume} {03}},\ \bibinfo
  {pages} {077} (\bibinfo {year} {2014})},\ \Eprint
  {http://arxiv.org/abs/1312.5330} {arXiv:1312.5330 [hep-ph]} \BibitemShut
  {NoStop}%
%%CITATION = ARXIV:1312.5330;%%
\bibitem [{\citenamefont {Ferretti}(2016)}]{Ferretti:2016upr}%
  \BibitemOpen
  \bibfield  {author} {\bibinfo {author} {\bibfnamefont {G.}~\bibnamefont
  {Ferretti}},\ }\href {\doibase 10.1007/JHEP06(2016)107} {\bibfield  {journal}
  {\bibinfo  {journal} {JHEP}\ }\textbf {\bibinfo {volume} {06}},\ \bibinfo
  {pages} {107} (\bibinfo {year} {2016})},\ \Eprint
  {http://arxiv.org/abs/1604.06467} {arXiv:1604.06467 [hep-ph]} \BibitemShut
  {NoStop}%
%%CITATION = ARXIV:1604.06467;%%
\bibitem [{\citenamefont {Brower}\ \emph {et~al.}(2014)\citenamefont {Brower},
  \citenamefont {Hasenfratz}, \citenamefont {Rebbi}, \citenamefont {Weinberg},\
  and\ \citenamefont {Witzel}}]{Brower:2014ita}%
  \BibitemOpen
  \bibfield  {author} {\bibinfo {author} {\bibfnamefont {R.~C.}\ \bibnamefont
  {Brower}}, \bibinfo {author} {\bibfnamefont {A.}~\bibnamefont {Hasenfratz}},
  \bibinfo {author} {\bibfnamefont {C.}~\bibnamefont {Rebbi}}, \bibinfo
  {author} {\bibfnamefont {E.}~\bibnamefont {Weinberg}}, \ and\ \bibinfo
  {author} {\bibfnamefont {O.}~\bibnamefont {Witzel}},\ }\href@noop {}
  {\bibfield  {journal} {\bibinfo  {journal} {PoS}\ }\textbf {\bibinfo {volume}
  {LATTICE2014}},\ \bibinfo {pages} {254} (\bibinfo {year} {2014})},\ \Eprint
  {http://arxiv.org/abs/1411.3243} {arXiv:1411.3243 [hep-lat]} \BibitemShut
  {NoStop}%
%%CITATION = ARXIV:1411.3243;%%
\bibitem [{\citenamefont {Weinberg}\ \emph {et~al.}(2015)\citenamefont
  {Weinberg}, \citenamefont {Brower}, \citenamefont {Hasenfratz}, \citenamefont
  {Rebbi},\ and\ \citenamefont {Witzel}}]{Weinberg:2014ega}%
  \BibitemOpen
  \bibfield  {author} {\bibinfo {author} {\bibfnamefont {E.}~\bibnamefont
  {Weinberg}}, \bibinfo {author} {\bibfnamefont {R.~C.}\ \bibnamefont
  {Brower}}, \bibinfo {author} {\bibfnamefont {A.}~\bibnamefont {Hasenfratz}},
  \bibinfo {author} {\bibfnamefont {C.}~\bibnamefont {Rebbi}}, \ and\ \bibinfo
  {author} {\bibfnamefont {O.}~\bibnamefont {Witzel}},\ }\href {\doibase
  10.1088/1742-6596/640/1/012055} {\bibfield  {journal} {\bibinfo  {journal}
  {J. Phys. Conf. Ser.}\ }\textbf {\bibinfo {volume} {640}},\ \bibinfo {pages}
  {012055} (\bibinfo {year} {2015})},\ \Eprint {http://arxiv.org/abs/1412.2148}
  {arXiv:1412.2148 [hep-lat]} \BibitemShut {NoStop}%
%%CITATION = ARXIV:1412.2148;%%
\bibitem [{\citenamefont {Hasenfratz}\ \emph
  {et~al.}(2017{\natexlab{b}})\citenamefont {Hasenfratz}, \citenamefont
  {Brower}, \citenamefont {Rebbi}, \citenamefont {Weinberg},\ and\
  \citenamefont {Witzel}}]{Hasenfratz:2015xca}%
  \BibitemOpen
  \bibfield  {author} {\bibinfo {author} {\bibfnamefont {A.}~\bibnamefont
  {Hasenfratz}}, \bibinfo {author} {\bibfnamefont {R.}~\bibnamefont {Brower}},
  \bibinfo {author} {\bibfnamefont {C.}~\bibnamefont {Rebbi}}, \bibinfo
  {author} {\bibfnamefont {E.}~\bibnamefont {Weinberg}}, \ and\ \bibinfo
  {author} {\bibfnamefont {O.}~\bibnamefont {Witzel}},\ }\href {\doibase
  10.1142/S0217751X17470030} {\bibfield  {journal} {\bibinfo  {journal} {Int.
  J. Mod. Phys. A}\ }\textbf {\bibinfo {volume} {32}},\ \bibinfo {pages}
  {1747003} (\bibinfo {year} {2017}{\natexlab{b}})},\ \Eprint
  {http://arxiv.org/abs/1510.04635} {arXiv:1510.04635 [hep-lat]} \BibitemShut
  {NoStop}%
\bibitem [{\citenamefont {Hasenfratz}\ \emph {et~al.}(2016)\citenamefont
  {Hasenfratz}, \citenamefont {Rebbi},\ and\ \citenamefont
  {Witzel}}]{Hasenfratz:2016uar}%
  \BibitemOpen
  \bibfield  {author} {\bibinfo {author} {\bibfnamefont {A.}~\bibnamefont
  {Hasenfratz}}, \bibinfo {author} {\bibfnamefont {C.}~\bibnamefont {Rebbi}}, \
  and\ \bibinfo {author} {\bibfnamefont {O.}~\bibnamefont {Witzel}},\
  }\href@noop {} {\bibfield  {journal} {\bibinfo  {journal} {PoS}\ }\textbf
  {\bibinfo {volume} {LATTICE2016}},\ \bibinfo {pages} {226} (\bibinfo {year}
  {2016})},\ \Eprint {http://arxiv.org/abs/1611.07427} {arXiv:1611.07427
  [hep-lat]} \BibitemShut {NoStop}%
%%CITATION = ARXIV:1611.07427;%%
\bibitem [{\citenamefont {Hasenfratz}\ \emph
  {et~al.}(2017{\natexlab{c}})\citenamefont {Hasenfratz}, \citenamefont
  {Rebbi},\ and\ \citenamefont {Witzel}}]{Hasenfratz:2017lne}%
  \BibitemOpen
  \bibfield  {author} {\bibinfo {author} {\bibfnamefont {A.}~\bibnamefont
  {Hasenfratz}}, \bibinfo {author} {\bibfnamefont {C.}~\bibnamefont {Rebbi}}, \
  and\ \bibinfo {author} {\bibfnamefont {O.}~\bibnamefont {Witzel}},\ }\href
  {\doibase 10.22323/1.314.0356} {\bibfield  {journal} {\bibinfo  {journal}
  {PoS}\ }\textbf {\bibinfo {volume} {EPS-HEP2017}},\ \bibinfo {pages} {356}
  (\bibinfo {year} {2017}{\natexlab{c}})},\ \Eprint
  {http://arxiv.org/abs/1710.02131} {arXiv:1710.02131 [hep-ph]} \BibitemShut
  {NoStop}%
%%CITATION = ARXIV:1710.02131;%%
\bibitem [{\citenamefont {Kaplan}(1992)}]{Kaplan:1992bt}%
  \BibitemOpen
  \bibfield  {author} {\bibinfo {author} {\bibfnamefont {D.~B.}\ \bibnamefont
  {Kaplan}},\ }\href {\doibase 10.1016/0370-2693(92)91112-M} {\bibfield
  {journal} {\bibinfo  {journal} {Phys. Lett.}\ }\textbf {\bibinfo {volume}
  {B288}},\ \bibinfo {pages} {342} (\bibinfo {year} {1992})},\ \Eprint
  {http://arxiv.org/abs/hep-lat/9206013} {arXiv:hep-lat/9206013} \BibitemShut
  {NoStop}%
%%CITATION = HEP-LAT/9206013;%%
\bibitem [{\citenamefont {Shamir}(1993)}]{Shamir:1993zy}%
  \BibitemOpen
  \bibfield  {author} {\bibinfo {author} {\bibfnamefont {Y.}~\bibnamefont
  {Shamir}},\ }\href {\doibase 10.1016/0550-3213(93)90162-I} {\bibfield
  {journal} {\bibinfo  {journal} {Nucl. Phys.}\ }\textbf {\bibinfo {volume}
  {B406}},\ \bibinfo {pages} {90} (\bibinfo {year} {1993})},\ \Eprint
  {http://arxiv.org/abs/hep-lat/9303005} {arXiv:hep-lat/9303005} \BibitemShut
  {NoStop}%
%%CITATION = HEP-LAT/9303005;%%
\bibitem [{\citenamefont {Furman}\ and\ \citenamefont
  {Shamir}(1995)}]{Furman:1994ky}%
  \BibitemOpen
  \bibfield  {author} {\bibinfo {author} {\bibfnamefont {V.}~\bibnamefont
  {Furman}}\ and\ \bibinfo {author} {\bibfnamefont {Y.}~\bibnamefont
  {Shamir}},\ }\href {\doibase 10.1016/0550-3213(95)00031-M} {\bibfield
  {journal} {\bibinfo  {journal} {Nucl. Phys.}\ }\textbf {\bibinfo {volume}
  {B439}},\ \bibinfo {pages} {54} (\bibinfo {year} {1995})},\ \Eprint
  {http://arxiv.org/abs/hep-lat/9405004} {arXiv:hep-lat/9405004} \BibitemShut
  {NoStop}%
%%CITATION = HEP-LAT/9405004;%%
\bibitem [{\citenamefont {Brower}\ \emph {et~al.}(2017)\citenamefont {Brower},
  \citenamefont {Neff},\ and\ \citenamefont {Orginos}}]{Brower:2012vk}%
  \BibitemOpen
  \bibfield  {author} {\bibinfo {author} {\bibfnamefont {R.~C.}\ \bibnamefont
  {Brower}}, \bibinfo {author} {\bibfnamefont {H.}~\bibnamefont {Neff}}, \ and\
  \bibinfo {author} {\bibfnamefont {K.}~\bibnamefont {Orginos}},\ }\href
  {\doibase 10.1016/j.cpc.2017.01.024} {\bibfield  {journal} {\bibinfo
  {journal} {Comput. Phys. Commun.}\ }\textbf {\bibinfo {volume} {220}},\
  \bibinfo {pages} {1} (\bibinfo {year} {2017})},\ \Eprint
  {http://arxiv.org/abs/1206.5214} {arXiv:1206.5214 [hep-lat]} \BibitemShut
  {NoStop}%
%%CITATION = ARXIV:1206.5214;%%
\bibitem [{\citenamefont {Witzel}\ \emph {et~al.}(2018)\citenamefont {Witzel},
  \citenamefont {Hasenfratz},\ and\ \citenamefont {Rebbi}}]{Witzel:2018gxm}%
  \BibitemOpen
  \bibfield  {author} {\bibinfo {author} {\bibfnamefont {O.}~\bibnamefont
  {Witzel}}, \bibinfo {author} {\bibfnamefont {A.}~\bibnamefont {Hasenfratz}},
  \ and\ \bibinfo {author} {\bibfnamefont {C.}~\bibnamefont {Rebbi}},\
  }\href@noop {} {\  (\bibinfo {year} {2018})},\ \Eprint
  {http://arxiv.org/abs/1810.01850} {arXiv:1810.01850 [hep-ph]} \BibitemShut
  {NoStop}%
%%CITATION = ARXIV:1810.01850;%%
\bibitem [{\citenamefont {Witzel}\ and\ \citenamefont
  {Hasenfratz}(2019)}]{Witzel:2019oej}%
  \BibitemOpen
  \bibfield  {author} {\bibinfo {author} {\bibfnamefont {O.}~\bibnamefont
  {Witzel}}\ and\ \bibinfo {author} {\bibfnamefont {A.}~\bibnamefont
  {Hasenfratz}} (\bibinfo {collaboration} {Lattice Strong Dynamics}),\ }\href
  {\doibase 10.22323/1.363.0115} {\bibfield  {journal} {\bibinfo  {journal}
  {PoS}\ }\textbf {\bibinfo {volume} {LATTICE2019}},\ \bibinfo {pages} {115}
  (\bibinfo {year} {2019})},\ \Eprint {http://arxiv.org/abs/1912.12255}
  {arXiv:1912.12255 [hep-lat]} \BibitemShut {NoStop}%
\bibitem [{\citenamefont {L{\"u}scher}\ and\ \citenamefont
  {Weisz}(1985{\natexlab{a}})}]{Luscher:1984xn}%
  \BibitemOpen
  \bibfield  {author} {\bibinfo {author} {\bibfnamefont {M.}~\bibnamefont
  {L{\"u}scher}}\ and\ \bibinfo {author} {\bibfnamefont {P.}~\bibnamefont
  {Weisz}},\ }\href {\doibase 10.1007/BF01206178} {\bibfield  {journal}
  {\bibinfo  {journal} {Commun. Math. Phys.}\ }\textbf {\bibinfo {volume}
  {97}},\ \bibinfo {pages} {59} (\bibinfo {year} {1985}{\natexlab{a}})},\
  \bibinfo {note} {[Erratum: Commun. Math. Phys.98,433(1985)]}\BibitemShut
  {NoStop}%
%%CITATION = CMPHA,97,59;%%
\bibitem [{\citenamefont {L{\"u}scher}\ and\ \citenamefont
  {Weisz}(1985{\natexlab{b}})}]{Luscher:1985zq}%
  \BibitemOpen
  \bibfield  {author} {\bibinfo {author} {\bibfnamefont {M.}~\bibnamefont
  {L{\"u}scher}}\ and\ \bibinfo {author} {\bibfnamefont {P.}~\bibnamefont
  {Weisz}},\ }\href {\doibase 10.1016/0370-2693(85)90966-9} {\bibfield
  {journal} {\bibinfo  {journal} {Phys. Lett.}\ }\textbf {\bibinfo {volume}
  {158B}},\ \bibinfo {pages} {250} (\bibinfo {year}
  {1985}{\natexlab{b}})}\BibitemShut {NoStop}%
%%CITATION = PHLTA,158B,250;%%
\bibitem [{\citenamefont {Morningstar}\ and\ \citenamefont
  {Peardon}(2004)}]{Morningstar:2003gk}%
  \BibitemOpen
  \bibfield  {author} {\bibinfo {author} {\bibfnamefont {C.}~\bibnamefont
  {Morningstar}}\ and\ \bibinfo {author} {\bibfnamefont {M.~J.}\ \bibnamefont
  {Peardon}},\ }\href {\doibase 10.1103/PhysRevD.69.054501} {\bibfield
  {journal} {\bibinfo  {journal} {Phys. Rev.}\ }\textbf {\bibinfo {volume}
  {D69}},\ \bibinfo {pages} {054501} (\bibinfo {year} {2004})},\ \Eprint
  {http://arxiv.org/abs/hep-lat/0311018} {arXiv:hep-lat/0311018 [hep-lat]}
  \BibitemShut {NoStop}%
%%CITATION = HEP-LAT/0311018;%%
\bibitem [{\citenamefont {Duane}\ \emph {et~al.}(1987)\citenamefont {Duane},
  \citenamefont {Kennedy}, \citenamefont {Pendleton},\ and\ \citenamefont
  {Roweth}}]{Duane:1987de}%
  \BibitemOpen
  \bibfield  {author} {\bibinfo {author} {\bibfnamefont {S.}~\bibnamefont
  {Duane}}, \bibinfo {author} {\bibfnamefont {A.}~\bibnamefont {Kennedy}},
  \bibinfo {author} {\bibfnamefont {B.}~\bibnamefont {Pendleton}}, \ and\
  \bibinfo {author} {\bibfnamefont {D.}~\bibnamefont {Roweth}},\ }\href
  {\doibase 10.1016/0370-2693(87)91197-X} {\bibfield  {journal} {\bibinfo
  {journal} {Phys.Lett.}\ }\textbf {\bibinfo {volume} {B195}},\ \bibinfo
  {pages} {216} (\bibinfo {year} {1987})}\BibitemShut {NoStop}%
%%CITATION = PHLTA,B195,216;%%
\bibitem [{\citenamefont {Wolff}(2004)}]{Wolff:2003sm}%
  \BibitemOpen
  \bibfield  {author} {\bibinfo {author} {\bibfnamefont {U.}~\bibnamefont
  {Wolff}} (\bibinfo {collaboration} {ALPHA}),\ }\href {\doibase
  10.1016/S0010-4655(03)00467-3, 10.1016/j.cpc.2006.12.001} {\bibfield
  {journal} {\bibinfo  {journal} {Comput.Phys.Commun.}\ }\textbf {\bibinfo
  {volume} {156}},\ \bibinfo {pages} {143} (\bibinfo {year} {2004})},\ \Eprint
  {http://arxiv.org/abs/hep-lat/0306017} {arXiv:hep-lat/0306017 [hep-lat]}
  \BibitemShut {NoStop}%
%%CITATION = HEP-LAT/0306017;%%
\bibitem [{\citenamefont {DeGrand}\ and\ \citenamefont
  {Hasenfratz}(2009)}]{DeGrand:2009mt}%
  \BibitemOpen
  \bibfield  {author} {\bibinfo {author} {\bibfnamefont {T.}~\bibnamefont
  {DeGrand}}\ and\ \bibinfo {author} {\bibfnamefont {A.}~\bibnamefont
  {Hasenfratz}},\ }\href {\doibase 10.1103/PhysRevD.80.034506} {\bibfield
  {journal} {\bibinfo  {journal} {Phys. Rev.}\ }\textbf {\bibinfo {volume}
  {D80}},\ \bibinfo {pages} {034506} (\bibinfo {year} {2009})},\ \Eprint
  {http://arxiv.org/abs/0906.1976} {arXiv:0906.1976 [hep-lat]} \BibitemShut
  {NoStop}%
%%CITATION = ARXIV:0906.1976;%%
\bibitem [{\citenamefont {Del~Debbio}\ and\ \citenamefont
  {Zwicky}(2010)}]{DelDebbio:2010ze}%
  \BibitemOpen
  \bibfield  {author} {\bibinfo {author} {\bibfnamefont {L.}~\bibnamefont
  {Del~Debbio}}\ and\ \bibinfo {author} {\bibfnamefont {R.}~\bibnamefont
  {Zwicky}},\ }\href {\doibase 10.1103/PhysRevD.82.014502} {\bibfield
  {journal} {\bibinfo  {journal} {Phys. Rev.}\ }\textbf {\bibinfo {volume}
  {D82}},\ \bibinfo {pages} {014502} (\bibinfo {year} {2010})},\ \Eprint
  {http://arxiv.org/abs/1005.2371} {arXiv:1005.2371 [hep-ph]} \BibitemShut
  {NoStop}%
%%CITATION = ARXIV:1005.2371;%%
\bibitem [{\citenamefont {Del~Debbio}\ and\ \citenamefont
  {Zwicky}(2011)}]{DelDebbio:2010jy}%
  \BibitemOpen
  \bibfield  {author} {\bibinfo {author} {\bibfnamefont {L.}~\bibnamefont
  {Del~Debbio}}\ and\ \bibinfo {author} {\bibfnamefont {R.}~\bibnamefont
  {Zwicky}},\ }\href {\doibase 10.1016/j.physletb.2011.04.059} {\bibfield
  {journal} {\bibinfo  {journal} {Phys. Lett. B}\ }\textbf {\bibinfo {volume}
  {700}},\ \bibinfo {pages} {217} (\bibinfo {year} {2011})},\ \Eprint
  {http://arxiv.org/abs/1009.2894} {arXiv:1009.2894 [hep-ph]} \BibitemShut
  {NoStop}%
\bibitem [{\citenamefont {L{\"{u}}scher}(2010)}]{Luscher:2010iy}%
  \BibitemOpen
  \bibfield  {author} {\bibinfo {author} {\bibfnamefont {M.}~\bibnamefont
  {L{\"{u}}scher}},\ }\href {\doibase 10.1007/JHEP08(2010)071} {\bibfield
  {journal} {\bibinfo  {journal} {JHEP}\ }\textbf {\bibinfo {volume} {1008}},\
  \bibinfo {pages} {071} (\bibinfo {year} {2010})},\ \Eprint
  {http://arxiv.org/abs/1006.4518} {arXiv:1006.4518 [hep-lat]} \BibitemShut
  {NoStop}%
%%CITATION = ARXIV:1006.4518;%%
\bibitem [{\citenamefont {Hasenfratz}\ and\ \citenamefont
  {Witzel}(tion)}]{AnomDim:2020}%
  \BibitemOpen
  \bibfield  {author} {\bibinfo {author} {\bibfnamefont {A.}~\bibnamefont
  {Hasenfratz}}\ and\ \bibinfo {author} {\bibfnamefont {O.}~\bibnamefont
  {Witzel}},\ }\href@noop {} {\enquote {\bibinfo {title} {{The running
  anomalous dimension of fermion bi- and trilinear operators in the 10-flavor
  SU(3) system}},}\ } (\bibinfo {year} {2020, in preparation}),\ \bibinfo
  {note} {in preparation}\BibitemShut {NoStop}%
\bibitem [{\citenamefont {Ryttov}\ and\ \citenamefont
  {Shrock}(2016{\natexlab{c}})}]{Ryttov:2016asb}%
  \BibitemOpen
  \bibfield  {author} {\bibinfo {author} {\bibfnamefont {T.~A.}\ \bibnamefont
  {Ryttov}}\ and\ \bibinfo {author} {\bibfnamefont {R.}~\bibnamefont
  {Shrock}},\ }\href {\doibase 10.1103/PhysRevD.94.105014} {\bibfield
  {journal} {\bibinfo  {journal} {Phys. Rev. D}\ }\textbf {\bibinfo {volume}
  {94}},\ \bibinfo {pages} {105014} (\bibinfo {year} {2016}{\natexlab{c}})},\
  \Eprint {http://arxiv.org/abs/1608.00068} {arXiv:1608.00068 [hep-th]}
  \BibitemShut {NoStop}%
\bibitem [{\citenamefont {Ryttov}\ and\ \citenamefont
  {Shrock}(2018)}]{Ryttov:2017lkz}%
  \BibitemOpen
  \bibfield  {author} {\bibinfo {author} {\bibfnamefont {T.~A.}\ \bibnamefont
  {Ryttov}}\ and\ \bibinfo {author} {\bibfnamefont {R.}~\bibnamefont
  {Shrock}},\ }\href {\doibase 10.1103/PhysRevD.97.025004} {\bibfield
  {journal} {\bibinfo  {journal} {Phys. Rev. D}\ }\textbf {\bibinfo {volume}
  {97}},\ \bibinfo {pages} {025004} (\bibinfo {year} {2018})},\ \Eprint
  {http://arxiv.org/abs/1710.06944} {arXiv:1710.06944 [hep-th]} \BibitemShut
  {NoStop}%
\bibitem [{\citenamefont {Matsuzaki}\ and\ \citenamefont
  {Yamawaki}(2014)}]{Matsuzaki:2013eva}%
  \BibitemOpen
  \bibfield  {author} {\bibinfo {author} {\bibfnamefont {S.}~\bibnamefont
  {Matsuzaki}}\ and\ \bibinfo {author} {\bibfnamefont {K.}~\bibnamefont
  {Yamawaki}},\ }\href {\doibase 10.1103/PhysRevLett.113.082002} {\bibfield
  {journal} {\bibinfo  {journal} {Phys. Rev. Lett.}\ }\textbf {\bibinfo
  {volume} {113}},\ \bibinfo {pages} {082002} (\bibinfo {year} {2014})},\
  \Eprint {http://arxiv.org/abs/1311.3784} {arXiv:1311.3784 [hep-lat]}
  \BibitemShut {NoStop}%
\bibitem [{\citenamefont {Appelquist}\ \emph {et~al.}(2016)\citenamefont
  {Appelquist}, \citenamefont {Brower}, \citenamefont {Fleming}, \citenamefont
  {Hasenfratz}, \citenamefont {Jin}, \citenamefont {Kiskis}, \citenamefont
  {Neil}, \citenamefont {Osborn}, \citenamefont {Rebbi}, \citenamefont
  {Rinaldi}, \citenamefont {Schaich}, \citenamefont {Vranas}, \citenamefont
  {Weinberg},\ and\ \citenamefont {Witzel}}]{Appelquist:2016viq}%
  \BibitemOpen
  \bibfield  {author} {\bibinfo {author} {\bibfnamefont {T.}~\bibnamefont
  {Appelquist}}, \bibinfo {author} {\bibfnamefont {R.}~\bibnamefont {Brower}},
  \bibinfo {author} {\bibfnamefont {G.}~\bibnamefont {Fleming}}, \bibinfo
  {author} {\bibfnamefont {A.}~\bibnamefont {Hasenfratz}}, \bibinfo {author}
  {\bibfnamefont {X.-Y.}\ \bibnamefont {Jin}}, \bibinfo {author} {\bibfnamefont
  {J.}~\bibnamefont {Kiskis}}, \bibinfo {author} {\bibfnamefont
  {E.}~\bibnamefont {Neil}}, \bibinfo {author} {\bibfnamefont {J.}~\bibnamefont
  {Osborn}}, \bibinfo {author} {\bibfnamefont {C.}~\bibnamefont {Rebbi}},
  \bibinfo {author} {\bibfnamefont {E.}~\bibnamefont {Rinaldi}}, \bibinfo
  {author} {\bibfnamefont {D.}~\bibnamefont {Schaich}}, \bibinfo {author}
  {\bibfnamefont {P.}~\bibnamefont {Vranas}}, \bibinfo {author} {\bibfnamefont
  {E.}~\bibnamefont {Weinberg}}, \ and\ \bibinfo {author} {\bibfnamefont
  {O.}~\bibnamefont {Witzel}} (\bibinfo {collaboration} {Lattice Strong
  Dynamics}),\ }\href {\doibase 10.1103/PhysRevD.93.114514} {\bibfield
  {journal} {\bibinfo  {journal} {Phys. Rev.}\ }\textbf {\bibinfo {volume}
  {D93}},\ \bibinfo {pages} {114514} (\bibinfo {year} {2016})},\ \Eprint
  {http://arxiv.org/abs/1601.04027} {arXiv:1601.04027 [hep-lat]} \BibitemShut
  {NoStop}%
%%CITATION = ARXIV:1601.04027;%%
\bibitem [{\citenamefont {Appelquist}\ \emph {et~al.}(2019)\citenamefont
  {Appelquist}, \citenamefont {Brower}, \citenamefont {Fleming}, \citenamefont
  {Gasbarro}, \citenamefont {Hasenfratz}, \citenamefont {Jin}, \citenamefont
  {Neil}, \citenamefont {Osborn}, \citenamefont {Rebbi}, \citenamefont
  {Rinaldi}, \citenamefont {Schaich}, \citenamefont {Vranas}, \citenamefont
  {Weinberg},\ and\ \citenamefont {Witzel}}]{Appelquist:2018yqe}%
  \BibitemOpen
  \bibfield  {author} {\bibinfo {author} {\bibfnamefont {T.}~\bibnamefont
  {Appelquist}}, \bibinfo {author} {\bibfnamefont {R.}~\bibnamefont {Brower}},
  \bibinfo {author} {\bibfnamefont {G.}~\bibnamefont {Fleming}}, \bibinfo
  {author} {\bibfnamefont {A.}~\bibnamefont {Gasbarro}}, \bibinfo {author}
  {\bibfnamefont {A.}~\bibnamefont {Hasenfratz}}, \bibinfo {author}
  {\bibfnamefont {X.-Y.}\ \bibnamefont {Jin}}, \bibinfo {author} {\bibfnamefont
  {E.}~\bibnamefont {Neil}}, \bibinfo {author} {\bibfnamefont {J.}~\bibnamefont
  {Osborn}}, \bibinfo {author} {\bibfnamefont {C.}~\bibnamefont {Rebbi}},
  \bibinfo {author} {\bibfnamefont {E.}~\bibnamefont {Rinaldi}}, \bibinfo
  {author} {\bibfnamefont {D.}~\bibnamefont {Schaich}}, \bibinfo {author}
  {\bibfnamefont {P.}~\bibnamefont {Vranas}}, \bibinfo {author} {\bibfnamefont
  {E.}~\bibnamefont {Weinberg}}, \ and\ \bibinfo {author} {\bibfnamefont
  {O.}~\bibnamefont {Witzel}} (\bibinfo {collaboration} {Lattice Strong
  Dynamics}),\ }\href {\doibase 10.1103/PhysRevD.99.014509} {\bibfield
  {journal} {\bibinfo  {journal} {Phys. Rev.}\ }\textbf {\bibinfo {volume}
  {D99}},\ \bibinfo {pages} {014509} (\bibinfo {year} {2019})},\ \Eprint
  {http://arxiv.org/abs/1807.08411} {arXiv:1807.08411 [hep-lat]} \BibitemShut
  {NoStop}%
%%CITATION = ARXIV:1807.08411;%%
\bibitem [{\citenamefont {Appelquist}\ \emph {et~al.}(2011)\citenamefont
  {Appelquist}, \citenamefont {Fleming}, \citenamefont {Lin}, \citenamefont
  {Neil},\ and\ \citenamefont {Schaich}}]{Appelquist:2011dp}%
  \BibitemOpen
  \bibfield  {author} {\bibinfo {author} {\bibfnamefont {T.}~\bibnamefont
  {Appelquist}}, \bibinfo {author} {\bibfnamefont {G.}~\bibnamefont {Fleming}},
  \bibinfo {author} {\bibfnamefont {M.}~\bibnamefont {Lin}}, \bibinfo {author}
  {\bibfnamefont {E.}~\bibnamefont {Neil}}, \ and\ \bibinfo {author}
  {\bibfnamefont {D.}~\bibnamefont {Schaich}},\ }\href {\doibase
  10.1103/PhysRevD.84.054501} {\bibfield  {journal} {\bibinfo  {journal}
  {Phys.Rev.}\ }\textbf {\bibinfo {volume} {D84}},\ \bibinfo {pages} {054501}
  (\bibinfo {year} {2011})},\ \Eprint {http://arxiv.org/abs/1106.2148}
  {arXiv:1106.2148 [hep-lat]} \BibitemShut {NoStop}%
%%CITATION = ARXIV:1106.2148;%%
\bibitem [{\citenamefont {DeGrand}(2011)}]{DeGrand:2011cu}%
  \BibitemOpen
  \bibfield  {author} {\bibinfo {author} {\bibfnamefont {T.}~\bibnamefont
  {DeGrand}},\ }\href {\doibase 10.1103/PhysRevD.84.116901} {\bibfield
  {journal} {\bibinfo  {journal} {Phys.Rev.}\ }\textbf {\bibinfo {volume}
  {D84}},\ \bibinfo {pages} {116901} (\bibinfo {year} {2011})},\ \Eprint
  {http://arxiv.org/abs/1109.1237} {arXiv:1109.1237 [hep-lat]} \BibitemShut
  {NoStop}%
%%CITATION = ARXIV:1109.1237;%%
\bibitem [{\citenamefont {Cheng}\ \emph {et~al.}(2013)\citenamefont {Cheng},
  \citenamefont {Hasenfratz}, \citenamefont {Petropoulos},\ and\ \citenamefont
  {Schaich}}]{Cheng:2013eu}%
  \BibitemOpen
  \bibfield  {author} {\bibinfo {author} {\bibfnamefont {A.}~\bibnamefont
  {Cheng}}, \bibinfo {author} {\bibfnamefont {A.}~\bibnamefont {Hasenfratz}},
  \bibinfo {author} {\bibfnamefont {G.}~\bibnamefont {Petropoulos}}, \ and\
  \bibinfo {author} {\bibfnamefont {D.}~\bibnamefont {Schaich}},\ }\href
  {\doibase 10.1007/JHEP07(2013)061} {\bibfield  {journal} {\bibinfo  {journal}
  {JHEP}\ }\textbf {\bibinfo {volume} {1307}},\ \bibinfo {pages} {061}
  (\bibinfo {year} {2013})},\ \Eprint {http://arxiv.org/abs/1301.1355}
  {arXiv:1301.1355 [hep-lat]} \BibitemShut {NoStop}%
%%CITATION = ARXIV:1301.1355;%%
\bibitem [{\citenamefont {Cheng}\ \emph {et~al.}(2014)\citenamefont {Cheng},
  \citenamefont {Hasenfratz}, \citenamefont {Liu}, \citenamefont
  {Petropoulos},\ and\ \citenamefont {Schaich}}]{Cheng:2013xha}%
  \BibitemOpen
  \bibfield  {author} {\bibinfo {author} {\bibfnamefont {A.}~\bibnamefont
  {Cheng}}, \bibinfo {author} {\bibfnamefont {A.}~\bibnamefont {Hasenfratz}},
  \bibinfo {author} {\bibfnamefont {Y.}~\bibnamefont {Liu}}, \bibinfo {author}
  {\bibfnamefont {G.}~\bibnamefont {Petropoulos}}, \ and\ \bibinfo {author}
  {\bibfnamefont {D.}~\bibnamefont {Schaich}},\ }\href {\doibase
  10.1103/PhysRevD.90.014509} {\bibfield  {journal} {\bibinfo  {journal}
  {Phys.Rev.}\ }\textbf {\bibinfo {volume} {D90}},\ \bibinfo {pages} {014509}
  (\bibinfo {year} {2014})},\ \Eprint {http://arxiv.org/abs/1401.0195}
  {arXiv:1401.0195 [hep-lat]} \BibitemShut {NoStop}%
%%CITATION = ARXIV:1401.0195;%%
\bibitem [{\citenamefont {Lombardo}\ \emph {et~al.}(2014)\citenamefont
  {Lombardo}, \citenamefont {Miura}, \citenamefont {da~Silva},\ and\
  \citenamefont {Pallante}}]{Lombardo:2014pda}%
  \BibitemOpen
  \bibfield  {author} {\bibinfo {author} {\bibfnamefont {M.~P.}\ \bibnamefont
  {Lombardo}}, \bibinfo {author} {\bibfnamefont {K.}~\bibnamefont {Miura}},
  \bibinfo {author} {\bibfnamefont {T.~J.~N.}\ \bibnamefont {da~Silva}}, \ and\
  \bibinfo {author} {\bibfnamefont {E.}~\bibnamefont {Pallante}},\ }\href
  {\doibase 10.1007/JHEP12(2014)183} {\bibfield  {journal} {\bibinfo  {journal}
  {JHEP}\ }\textbf {\bibinfo {volume} {12}},\ \bibinfo {pages} {183} (\bibinfo
  {year} {2014})},\ \Eprint {http://arxiv.org/abs/1410.0298} {arXiv:1410.0298
  [hep-lat]} \BibitemShut {NoStop}%
%%CITATION = ARXIV:1410.0298;%%
\bibitem [{\citenamefont {Li}\ and\ \citenamefont {Poland}(2020)}]{Li:2020bnb}%
  \BibitemOpen
  \bibfield  {author} {\bibinfo {author} {\bibfnamefont {Z.}~\bibnamefont
  {Li}}\ and\ \bibinfo {author} {\bibfnamefont {D.}~\bibnamefont {Poland}},\
  }\href@noop {} {\  (\bibinfo {year} {2020})},\ \Eprint
  {http://arxiv.org/abs/2005.01721} {arXiv:2005.01721 [hep-th]} \BibitemShut
  {NoStop}%
\bibitem [{\citenamefont {Bär}\ and\ \citenamefont
  {Golterman}(2014)}]{Bar:2013ora}%
  \BibitemOpen
  \bibfield  {author} {\bibinfo {author} {\bibfnamefont {O.}~\bibnamefont
  {Bär}}\ and\ \bibinfo {author} {\bibfnamefont {M.}~\bibnamefont
  {Golterman}},\ }\href {\doibase 10.1103/PhysRevD.89.034505} {\bibfield
  {journal} {\bibinfo  {journal} {Phys. Rev. D}\ }\textbf {\bibinfo {volume}
  {89}},\ \bibinfo {pages} {034505} (\bibinfo {year} {2014})},\ \bibinfo {note}
  {[Erratum: Phys.Rev.D 89, 099905 (2014)]},\ \Eprint
  {http://arxiv.org/abs/1312.4999} {arXiv:1312.4999 [hep-lat]} \BibitemShut
  {NoStop}%
\bibitem [{\citenamefont
  {\href{https://github.com/paboyle/Grid}{https://github.com/paboyle/Grid}}()}]{Gridurl}%
  \BibitemOpen
  \bibfield  {author} {\bibinfo {author} {\bibnamefont
  {\href{https://github.com/paboyle/Grid}{https://github.com/paboyle/Grid}}},\
  }\href@noop {} {}\BibitemShut {NoStop}%
\bibitem [{\citenamefont {Boyle}\ \emph {et~al.}(2015)\citenamefont {Boyle},
  \citenamefont {Yamaguchi}, \citenamefont {Cossu},\ and\ \citenamefont
  {Portelli}}]{Boyle:2015tjk}%
  \BibitemOpen
  \bibfield  {author} {\bibinfo {author} {\bibfnamefont {P.}~\bibnamefont
  {Boyle}}, \bibinfo {author} {\bibfnamefont {A.}~\bibnamefont {Yamaguchi}},
  \bibinfo {author} {\bibfnamefont {G.}~\bibnamefont {Cossu}}, \ and\ \bibinfo
  {author} {\bibfnamefont {A.}~\bibnamefont {Portelli}},\ }\href {\doibase
  10.22323/1.251.0023} {\bibfield  {journal} {\bibinfo  {journal} {PoS}\
  }\textbf {\bibinfo {volume} {LATTICE2015}},\ \bibinfo {pages} {023} (\bibinfo
  {year} {2015})},\ \Eprint {http://arxiv.org/abs/1512.03487} {arXiv:1512.03487
  [hep-lat]} \BibitemShut {NoStop}%
%%CITATION = ARXIV:1512.03487;%%
\bibitem [{\citenamefont
  {\href{https://usqcd.lns.mit.edu/w/index.php/QLUA}{https://usqcd.lns.mit.edu/w/index.php/QLUA}}()}]{QLUAurl}%
  \BibitemOpen
  \bibfield  {author} {\bibinfo {author} {\bibnamefont
  {\href{https://usqcd.lns.mit.edu/w/index.php/QLUA}{https://usqcd.lns.mit.edu/w/index.php/QLUA}}},\
  }\href@noop {} {}\BibitemShut {NoStop}%
\bibitem [{\citenamefont {Pochinsky}(2008)}]{Pochinsky:2008zz}%
  \BibitemOpen
  \bibfield  {author} {\bibinfo {author} {\bibfnamefont {A.}~\bibnamefont
  {Pochinsky}},\ }\href {\doibase 10.22323/1.066.0040} {\bibfield  {journal}
  {\bibinfo  {journal} {PoS}\ }\textbf {\bibinfo {volume} {LATTICE2008}},\
  \bibinfo {pages} {040} (\bibinfo {year} {2008})}\BibitemShut {NoStop}%
%%CITATION = POSCI,LATTICE2008,040;%%
\bibitem [{\citenamefont {Anderson}\ \emph {et~al.}(2017)\citenamefont
  {Anderson}, \citenamefont {Burns}, \citenamefont {Milroy}, \citenamefont
  {Ruprecht}, \citenamefont {Hauser},\ and\ \citenamefont {Siegel}}]{UCsummit}%
  \BibitemOpen
  \bibfield  {author} {\bibinfo {author} {\bibfnamefont {J.}~\bibnamefont
  {Anderson}}, \bibinfo {author} {\bibfnamefont {P.~J.}\ \bibnamefont {Burns}},
  \bibinfo {author} {\bibfnamefont {D.}~\bibnamefont {Milroy}}, \bibinfo
  {author} {\bibfnamefont {P.}~\bibnamefont {Ruprecht}}, \bibinfo {author}
  {\bibfnamefont {T.}~\bibnamefont {Hauser}}, \ and\ \bibinfo {author}
  {\bibfnamefont {H.~J.}\ \bibnamefont {Siegel}},\ }\href {\doibase
  10.1145/3093338.3093379} {\bibfield  {journal} {\bibinfo  {journal}
  {Proceedings of PEARC17}\ }\textbf {\bibinfo {volume} {8}},\ \bibinfo {pages}
  {1} (\bibinfo {year} {2017})}\BibitemShut {NoStop}%
\end{thebibliography}%
\bibliographystyle{apsrev4-1}

\end{document}